%% file: main.tex
\documentclass[sigconf,authorversion]{acmart}

\input{fronts/definitions.tex}

\input{fronts/preamble.tex}
\begin{document}

\input{fronts/title.tex}
\input{fronts/authors.tex}
\input{fronts/abstract.tex}
\input{fronts/ccs.tex}
\input{fronts/keyword.tex}

\maketitle 

\input{sections/introduction.tex}
\input{sections/related-work.tex}

\input{sections/formative-study.tex}

\input{sections/preference-collection-system.tex}

\input{sections/evaluation.tex}

\input{sections/result.tex}

\input{sections/discussion.tex}
\input{sections/conclusion.tex}
\input{sections/acknowledgement.tex}

\bibliographystyle{ACM-Reference-Format}
\bibliography{reference}

\newpage
\appendix
\input{sections/appendix/feature-definitions.tex}

\end{document}

%% file: fronts/definitions.tex
\usepackage{multirow}
\usepackage{subcaption}
\usepackage{cleveref}
\usepackage{array}
\usepackage{longtable}
\usepackage{enumitem}

\newlength{\colonewidth}
\newlength{\coltwowidth}
\newlength{\colthreewidth}
\newlength{\tabletextwidth}

\newcommand{\eg}{{e.g.,\ }}

\newcommand{\etc}{{etc.}}
\newcommand{\ie}{{i.e.,\ }}

\definecolor{oxfordblue}{rgb}{0.0, 0.13, 0.28}
\definecolor{harvardcrimson}{rgb}{0.79, 0.0, 0.09}
\definecolor{dartmouthgreen}{rgb}{0.05, 0.5, 0.06}
\definecolor{princetonorange}{rgb}{1.0, 0.56, 0.0}
\definecolor{yaleblue}{rgb}{0.06, 0.3, 0.57}
\definecolor{usccardinal}{rgb}{0.6, 0.0, 0.0}
\definecolor{uclablue}{rgb}{0.33, 0.41, 0.58}
\definecolor{msugreen}{rgb}{0.09, 0.27, 0.23}
\definecolor{cornellred}{rgb}{0.7, 0.11, 0.11}
\definecolor{pomegranate}{RGB}{192, 57, 43}
\definecolor{anti-pomegranate}{RGB}{43,178,192}
\definecolor{alizarin}{RGB}{231, 76, 60}
\definecolor{peter}{RGB}{52, 152, 219}
\definecolor{green}{RGB}{22, 160, 133}
\definecolor{anti-green}{RGB}{160,22,118}
\definecolor{turquoise}{RGB}{26, 188, 156}
\definecolor{pumpkin}{RGB}{211, 84, 0}
\definecolor{anti-pumpkin}{RGB}{0,22,211}
\definecolor{carrot}{RGB}{230, 126, 34}
\definecolor{wisteria}{RGB}{142, 68, 173}
\definecolor{anti-wisteria}{RGB}{99,173,68}
\definecolor{amethyst}{RGB}{155, 89, 182}
\definecolor{sunflower}{RGB}{241, 196, 15}
\definecolor{MAIMIAOLV}{RGB}{85, 187, 138}


%% file: fronts/preamble.tex
\AtBeginDocument{%
  }

\copyrightyear{2024}
\acmYear{2024}
\setcopyright{acmlicensed}\acmConference[IUI '24]{29th International Conference on Intelligent User Interfaces}{March 18--21, 2024}{Greenville, SC, USA}
\acmBooktitle{29th International Conference on Intelligent User Interfaces (IUI '24), March 18--21, 2024, Greenville, SC, USA}
\acmDOI{10.1145/3640543.3645145}
\acmISBN{979-8-4007-0508-3/24/03}

%% file: fronts/title.tex
\newcommand{\dataset}{\textit{PickPlaceCans}}
\newcommand{\tool}{FARPLS}
\title[\tool: Feature-Augmented Robot Trajectory Preference Labeling System]{{\tool: A Feature-Augmented Robot Trajectory Preference Labeling System to Assist Human Labelers' Preference Elicitation}}

%% file: fronts/authors.tex

\author{Hanfang Lyu}
\email{hanfang.lyu@connect.ust.hk}
\orcid{0000-0003-0135-5754}
\affiliation{%
    \institution{Hong Kong University of Science and Technology}
    \city{Hong Kong}
    \country{China}
}

\author{Yuanchen Bai}
\email{ybai2@andrew.cmu.edu}
\orcid{0009-0004-2140-7894}
\affiliation{%
    \institution{Carnegie Mellon University}
    \city{Pittsburgh}
    \country{United States}
}

\author{Xin Liang}
\email{XinLiang0920@gmail.com}
\orcid{0009-0003-1497-4404}
\affiliation{
    \institution{Tongji University}
    \city{Shanghai}
    \country{China}
}

\author{Ujaan Das}
\email{udas@connect.ust.hk}
\orcid{0009-0001-6852-3938}
\affiliation{
    \institution{Hong Kong University of Science and Technology}
    \city{Hong Kong}
    \country{China}
}

\author{Chuhan Shi}
\email{cshiag@connect.ust.hk}
\orcid{0000-0002-3370-1626}
\affiliation{
    \institution{Southeast University}
    \city{Nanjing}
    \country{China}
}

\author{Leiliang Gong}
\email{leiliang@hkpc.org}
\orcid{0000-0003-4302-1592}
\affiliation{
    \institution{Robotics and AI Division, Hong Kong Productivity Council}
    \city{Hong Kong}
    \country{China}
}

\author{Yingchi Li}
\email{terenceli@hkflair.org}
\orcid{0009-0003-9457-4303}
\affiliation{
    \institution{Hong Kong Industrial Artificial Intelligence and Robotics Centre (FLAIR)}
    \city{Hong Kong}
    \country{China}
}

\author{Mingfei Sun}
\email{mingfei.sun@manchester.ac.uk}
\orcid{0000-0002-5925-5425}
\affiliation{
    \institution{Department of Computer Science, University of Manchester}
    \city{Manchester}
    \country{United Kingdom}
}

\author{Ming Ge}
\email{mingge@hkpc.org; mingge@hkflair.org}
\orcid{0000-0002-2768-1374}
\affiliation{
    \institution{Hong Kong Productivity Council; Hong Kong Industrial Artificial Intelligence and Robotics Centre (FLAIR)}
    \city{Hong Kong}
    \country{China}
}

\author{Xiaojuan Ma}
\authornote{Corresponding author}
\email{mxj@cse.ust.hk}
\orcid{0000-0002-9847-7784}
\affiliation{%
    \institution{Hong Kong University of Science and Technology}
    \city{Hong Kong}
    \country{China}
}

\renewcommand{\shortauthors}{Lyu, et al.}

%% file: fronts/abstract.tex
\begin{abstract}
    
    Preference-based learning aims to align robot task objectives with human values. One of the most common methods to infer human preferences is by pairwise comparisons of robot task trajectories.
    Traditional comparison-based preference labeling systems seldom support labelers to digest and identify critical differences between complex trajectories recorded in videos.
    Our formative study (N = $12$) suggests that individuals may overlook non-salient task features and establish biased preference criteria during their preference elicitation process because of partial observations.
    In addition, they may experience mental fatigue when given many pairs to compare, causing their label quality to deteriorate.
    To mitigate these issues, we propose \tool, a \textbf{F}eature-\textbf{A}ugmented \textbf{R}obot trajectory \textbf{P}reference \textbf{L}abeling \textbf{S}ystem.
    \tool{} highlights potential outliers in a wide variety of task features that matter to humans and extracts the corresponding video keyframes for easy review and comparison.
    It also dynamically adjusts the labeling order according to users' familiarities, difficulties of the trajectory pair, and level of disagreements. At the same time, the system monitors labelers' consistency and provides feedback on labeling progress to keep labelers engaged.
    A between-subjects study (N = $42$, $105$ pairs of robot pick-and-place trajectories per person) shows that \tool{} can help users establish preference criteria more easily and notice more relevant details in the presented trajectories than the conventional interface.
    \tool{} also improves labeling consistency and engagement, mitigating challenges in preference elicitation without raising cognitive loads significantly. 
    

\end{abstract}

%% file: fronts/ccs.tex
\ccsdesc[500]{Human-centered computing~HCI design and evaluation methods}
\ccsdesc[500]{Human-centered computing~Interactive systems and tools}

%% file: fronts/keyword.tex
\keywords{data labeling, preference collection system, human-robot value alignment. }

%% file: sections/introduction.tex
\section{Introduction}
\label{sec:introduction}

Advancement in artificial intelligence (AI) and robotics technologies brings robots out of laboratories,
requiring them to perform daily tasks for or with humans
\cite{laplazaIVORobotNew2022, sandygulovaIndividualDifferencesChildren2022, muehlhausNeedThirdArm2023, mehtaArmADineUnderstandingDesign2018,mitterbergerInteractiveRoboticPlastering2022}.
To accommodate this requirement, robot task learning aims at teaching robot tasks such as
prehabilitation \cite{woodworthPreferenceLearningAssistive2018}, companion \cite{ritschelAdaptiveLinguisticStyle2019}, assembly tasks \cite{nemlekarTransferLearningHuman2023} according to user preferences.
Traditional robot learning algorithms rely on delicately handcrafted reward functions to guide robot behavior.
However, these delicate reward functions may not accurately reflect humans' true values \cite{christianoDeepReinforcementLearning2017} due to generalization errors \cite{mohri2018foundations}, task misspecifications \cite{casperOpenProblemsFundamental2023}, etc.
The human-robot value alignment can not only improve robot performances according to humans' preference but also avoid undesired robot behavior and even safety issues \cite{yuanSituBidirectionalHumanrobot2022a,bobuAligningRobotHuman2023,sannemanTransparentValueAlignment2023}.
Learning a reward model from human preferences hence emerges \cite{christianoDeepReinforcementLearning2017},
which leverages the computationally efficient and user-friendly pairwise comparison to collect human preferences \cite{bradleyRankAnalysisIncomplete1952,hullermeierLabelRankingLearning2008, holladayActiveComparisonBased2016a, casperOpenProblemsFundamental2023}.
However, this learning process still requires a substantial number of high-quality human preference inputs,
which inevitably contain subjective uncertainties and incur a huge cognitive labor cost for participants.
Reducing the cost and uncertainty in the human preference data collection process has been one of the focuses of research on developing robots for human use.

The human preference collection process includes recruiting adequate human labelers to give high-quality preference annotations to many robot task trajectories.
The annotation requires human labelers to understand every robot trajectory presented
and to specify which trajectory is better via comparison from their point of view.
Previous studies mainly handled the challenges in human label collection from two algorithmic directions.
One line of research proposed to work with human data with inconsistent qualities,
trying to incorporate human uncertainty into active reward learning \cite{holladayActiveComparisonBased2016a}.
The other line of work focused primarily on reducing the number of human labels needed,
for example, by prompting pairs with a high information gain \cite{biyikAskingEasyQuestions2020}.
These studies, however, largely overlooked possibilities of improving data quality and human engagement by assisting in labelers' sensemaking process during preference elicitation.
Due to the complexity of full episodes of robot trajectory data,
understanding and comparing nuanced characteristics of robot task processes and performance based on trajectory-recording videos may be difficult for labelers,
especially those without robotics expertise.
Consequently, human labelers may find the trajectories presented to them arbitrary or misleading \cite{habibianHereWhatVe2022a},
leading to inconsistent labeling and low-quality preference data.
Thus, supporting human labelers' efficient labeling from novel interaction and design is an important aspect besides algorithmic perspectives.

In this work, we argue that the human preference collection system for robot task learning should not only focus on the algorithms' perspective of reducing the cost and uncertainties
but also provide sufficient support to human labelers and improve the data quality from the source.
We aim to address the following three research questions:
\begin{itemize}
    \item[\textbf{RQ1}] How do human labelers compare robot task trajectories, and what are their challenges and needs when eliciting their trajectory preferences?
    \item[\textbf{RQ2}] How to design a preference collection system that can assist human labelers in trajectory sensemaking and preference elicitation?
    \item[\textbf{RQ3}] How does the proposed system improve data quality and human engagement and mitigate the challenges above?
\end{itemize}

In this paper, we gain an understanding of how human labelers usually compare two robot task trajectories through semi-structured interviews in a formative study with $12$ participants.
We derive a list of trajectory features humans would consider as preference elicitation criteria by combining findings from previous literature and the formative study.
We also identify three main challenges that human labelers face in the preference collection process: difficulty in forming criteria, overlooking trajectory details, and difficulty in maintaining focus, which align with data labeling challenges in other fields (\eg \cite{herouxConsumerProductLabel1988, fredrikssonDataLabelingEmpirical2020}).

Drawing on the human needs and derived design requirements,
we designed \tool{}, a \textbf{F}eature-\textbf{A}ugmented \textbf{R}obot trajectory \textbf{P}ref-erence \textbf{L}abeling \textbf{S}ystem to address the challenges.
We generate a dataset of robot trajectories of a typical pick-and-place task commonly seen in human environments
and automatically extracted the features relevant to human criteria from the trajectories.
We clustered the trajectories based on these features and designed a prompting strategy to initially present the trajectory pairs that are varied in features to facilitate criteria formation.
\tool{} is a web-based system that allows human labelers to compare two robot trajectories pairwise.
The system marks feature-based keyframes in the video for accessible replay and comparison and highlights features that defer the most from the mean. 
The system also provides real-time attention monitoring and feedback to help human labelers maintain attention and engagement.

We conducted a between-subjects study with $42$ participants to compare the proposed tool with the conventional pairwise comparison interface.
Each participant was required to label $105$ pairs of robot pick-and-place trajectories and fill out a post-study questionnaire regarding confidence, cognitive load, and challenges from the formative study.
The results show that \tool{} significantly improves the labeling consistency of human labelers without raising their cognitive loads.
The subjective ratings from the questionnaire also show that \tool{} is significantly better than the baseline in terms of establishing comparison criteria and providing sensemaking supports to help labelers notice more nuanced details.
Although the average labeling time of labelers using \tool{} is significantly longer than those with the baseline system, the participants' perceived engagement in rewards and boredom in the labeling process is significantly improved.

The key contributions of our work are as follows:
\begin{itemize}
    \item We conducted a formative study to understand human labelers' pairwise preference elicitation process and their needs and challenges.
    \item We propose a novel trajectory preference labeling system for robot manipulation tasks, \tool, that can help human labelers give high-quality preference data.
    \item We conducted a user study to evaluate the effectiveness of our system and provide design considerations for future data collection and reward learning systems.
\end{itemize}

%% file: sections/related-work.tex
\section{Related Work}
\label{sec:related-work}

\subsection{Human Preference Learning in Robot Manipulation Tasks}
\label{sec:related-work-human-preference-learning}
Recent research has pointed out the potential misalignment issue between human values and robotic objectives.
Studies to address this issue include bidirectional human-robot communication in group settings \cite{yuanSituBidirectionalHumanrobot2022a},
evaluation of task accomplishment \cite{bobuAligningRobotHuman2023}, and disentangled representation learning (DRL) \cite{wangDRAVAAligningHuman2023}.
Particularly, Reinforcement Learning from Human Preference (RLHP) \cite{zhanHumanGuidedRobotBehavior2021,abramsonImprovingMultimodalInteractive2022, liReinforcementLearningHuman2023,liuPerspectivesSocialImpacts2023,casperOpenProblemsFundamental2023}
emerges as a new trend to offer a flexible and adaptable way to fine-tune an agent's behavior based on human preference.
RLHP comes with three key steps: (1) human preference collection, (2) reward learning, and (3) RL policy optimization.
Human preference has been collected in various ways including 
absolute rating \cite{carteretteHereThere2008},
and ranking
(voting \cite{conitzerElicitingSinglepeakedPreferences2007},
pairwise comparison \cite{kuhlmanEvaluatingPreferenceCollection2019a, qianLearningUserPreferences2015, glickmanAdaptivePairedComparison2005},
multiple ranking \cite{brownExtrapolatingSuboptimalDemonstrations2019, brownSafeImitationLearning2020, zhuPrincipledReinforcementLearning2023a, myersLearningMultimodalRewards2022a}, \etc).
Among them, pairwise comparison is the most widely used due to its advantages in terms of its impacts on model performance \cite{koczkodajTestingAccuracyEnhancement1998, hullermeierLabelRankingLearning2008a} and labeler experience \cite{holladayActiveComparisonBased2016a, bradleyRankAnalysisIncomplete1952}. 

These existing studies on human preference collection and learning primarily focus on the trajectory querying strategy, human data augmentation and representation \cite{casperOpenProblemsFundamental2023}, with little attention paid to assisting the human preference labeling process,
especially on how human values can be aligned in robotic trajectory (\cite{aliasghariEffectsGazeArm2021,mummHumanrobotProxemicsPhysical2011}) for better labeling outcomes.
We focus on the robotic trajectory evaluation by exploring key features that are deemed valuable in the robotics community (\eg \cite{dobisEvaluationCriteriaTrajectories2022,zhaoFSWRobotSystem2021,wuDesignControlledSpatial1988,tanakaMinimumjerkTrajectoryGeneration2012,magidVoronoibasedTrajectoryOptimization2017}).
We present a system that conveys the feature information in a comprehensible manner to labelers with different levels of prior knowledge,
especially those non-expert labelers.
Further, our study identifies the multifaceted challenges of robotics trajectory preference labeling,
thus offering a chance to optimize the entire crucial procedure and its subsequent outcomes.

\subsection{Data Annotation Tools for Machine Learning}

Labeling tools have been designed to deal with various tasks and data types, such as text (AILA \cite{choiAILAAttentiveInteractive2019}, DUALIST \cite{settlesClosingLoopFast2011}), image (EasyAlbum \cite{cuiEasyAlbumInteractivePhoto2007a}, SAPHARI \cite{suhSemiautomaticPhotoAnnotation2007a}), video (MediaTable \cite{rooijMediaTableInteractiveCategorization2010a}, VoTT \cite{VoTT2023}), audios (VIA \cite{duttaAnnotationSoftwareImages2019}), and other special use cases like activity label from the elderly \cite{kimMyMoveFacilitatingOlder2022a}.
Three main branches of various techniques include semi-automatic labeling, active learning, and novel interaction and visual design \cite{zhangOneLabelerFlexibleSystem2022}.
Active learning (\eg ALVA \cite{kucherActiveLearningVisual2017}, DUALIST \cite{settlesClosingLoopFast2011}, MI3 \cite{zhangMI3MachineinitiatedIntelligent2021}) and semi-automatic labeling (\eg ISSE\cite{bryanISSEInteractiveSource2014a}, V-awake\cite{garciacaballeroVawake21stEurographics2019}) are widely applied to optimize the labeling process and outcomes from the algorithmic perspective.
Novel designs of the interfaces are essential in supporting efficient labeling, such as a combination of tabular and bucket list \cite{rooijMediaTableInteractiveCategorization2010a}, keywords highlighting \cite{choiAILAAttentiveInteractive2019}, and instances clustering \cite{cuiEasyAlbumInteractivePhoto2007a}.

Existing labeling tools are all designed for specific data types, and they fail to tackle the unique challenges of robotic trajectory labeling since robotic arm trajectory introduces unique complexities compared to the data types mentioned above.
The trajectory data are highly non-linear and non-stationary, and the human labelers may not have a clear understanding of their movements and effects on the environment.
We focus on understanding the trajectory labeling process and designing a labeling system to facilitate human labeling of robotic arm trajectories.

\subsection{Visualization in Human-Robot Interaction}

In the realm of human-robot collaboration, significant efforts have been directed toward enhancing the visualization and demonstration of robot tasks, 
as evidenced by previous research studies (\eg \cite{adamsCriticalConsiderationsHumanRobot2002a, chenHumanPerformanceIssues2007, crawlFiremapDynamicDataDriven2017}). 
To provide some more detailed examples: 
\citet{chandanLearningVisualizationPolicies2023a} develops an intelligent augmented reality (AR) agent that learns visualization policies aimed at enhancing efficiency and minimizing distractions for humans;  
\citet{draganLegibilityPredictabilityRobot2013a} enhances the interface for human communication with a virtual robot and improves the robot's knowledge representation through the use of 3D isometric visualization, along with providing the robot's first-person perspective; 
\citet{zhuVirtuallyAdaptedReality2017a} targets the challenge of visual mismatch with video capture, aiming to facilitate the overlay of visualizations onto video streams and extract the underlying algorithms utilized;
\citet{szafirConnectingHumanRobotInteraction2021} proposed a data-centric HRI framework and identified visualization design concepts to facilitate HRI data tasks such as data collection, analysis, and human decision-making.

Existing interfaces for HRI systems primarily focus on situation awareness and user control \cite{szafirConnectingHumanRobotInteraction2021, wangUserInterfaceSensemaking2023}. 
In these contexts, the interfaces focus on how to assist the human-robot interaction and data analysis process and fail to consider how to improve the human experience in the labeling process and the data quality.
It remains unclear what specific information is required and how it should be presented within the context of providing preferences for the robot arm manipulation tasks.
In our work, we will draw inspiration from previous interface design and data presentation approaches,
incorporating novel insights from a formative study to enhance the interaction between human labelers and our labeling system for robot arm trajectories.

%% file: sections/formative-study.tex
\section{Formative Study}
\label{sec:formative-study}

In this section, we present the formative study and findings to answer {RQ1}.
With the Institutional Review Board (IRB) approval from the {University} Research Ethics Committee,
we conducted a formative study to explore how labelers compare robot trajectories
and figure out challenges and needs in traditional pairwise trajectory preference collection systems.

\input{sections/formative-study/formative-study-design.tex}
\input{sections/formative-study/criteria-features.tex}
\input{sections/formative-study/challenges.tex}

\begin{figure*}
    \centering
    \includegraphics[width=\textwidth]{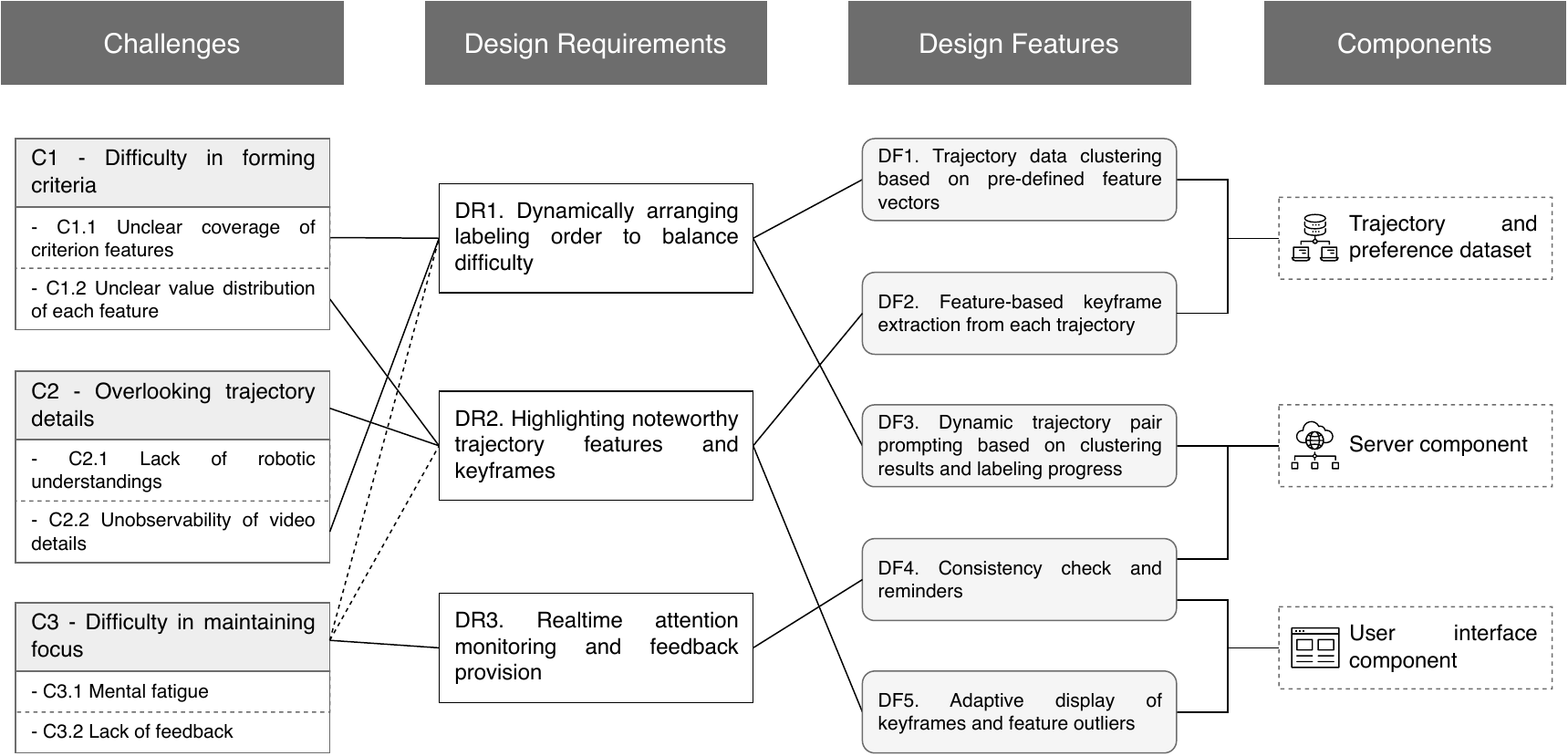}
    
    \caption{{This figure demonstrates an overview of our design pipeline: from identified challenges, we establish the design requirements to address each challenge, then design corresponding features to meet those requirements, and finally, we integrate those features into concrete components of our labeling system.}}
    \Description{Challenges, design requirements, design features, and components.}
    \label{fig:C-DR-DF}
\end{figure*}
\input{sections/formative-study/design-requirements.tex}

%% file: sections/formative-study/formative-study-design.tex
\subsection{Study Design and Procedure}
\label{sec:formative-study-design}

\subsubsection{Participants}
We recruited twelve participants ($6$ females and $6$ males) aged $19$ to $28$ ($M = 22.0$, $SD = 2.6$) to participate in the formative study. 
Their familiarity with the data labeling system ($1$ for Not familiar at all --- $5$ for Extremely familiar) ranges from $1$ to $3$ ($M = 1.9$, $SD = 0.9$). 
Additionally, their familiarity with robotics ($1$ for Not familiar at all --- $5$ for Extremely familiar) ranges from $1$ to $4$ ($M = 1.9$, $SD = 0.9$). 

\subsubsection{Materials}
We built a basic preference labeling system following the interface of \citet{christianoDeepReinforcementLearning2017} shown on their project website \footnote{\url{https://openai.com/research/learning-from-human-preferences}}.
We generated a pick-and-place dataset based on the \textit{Can} task of robomimic \footnote{\url{https://robomimic.github.io/docs/v0.2/datasets/robomimic_v0.1.html\#info}}, with six cans on the table and four bins on the side. 
\begin{figure*}
    \centering
    \begin{subfigure}{0.24\textwidth}
        \centering
        \includegraphics[width=\textwidth]{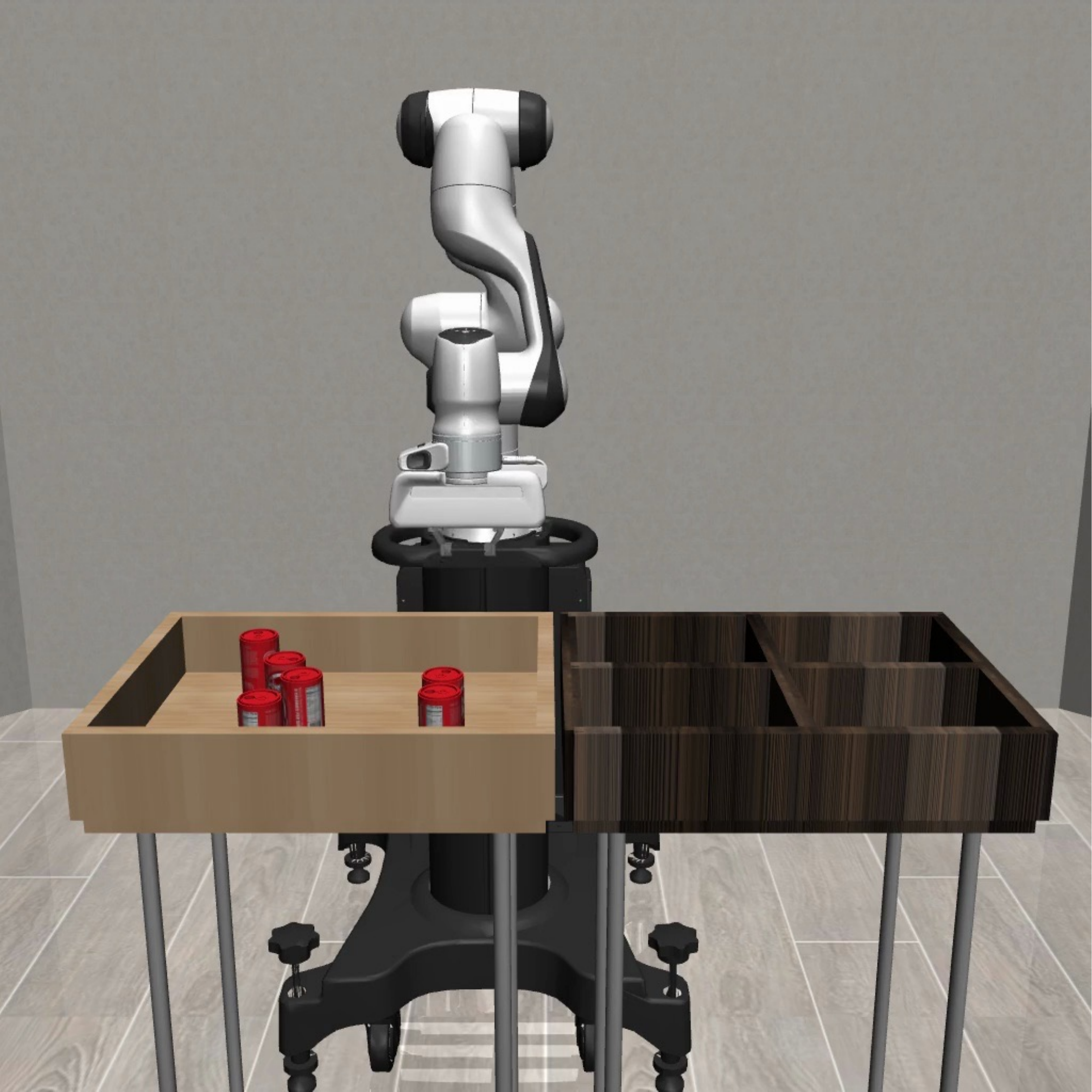}
    \end{subfigure}
    \hfil
    \begin{subfigure}{0.24\textwidth}
        \centering
        \includegraphics[width=\textwidth]{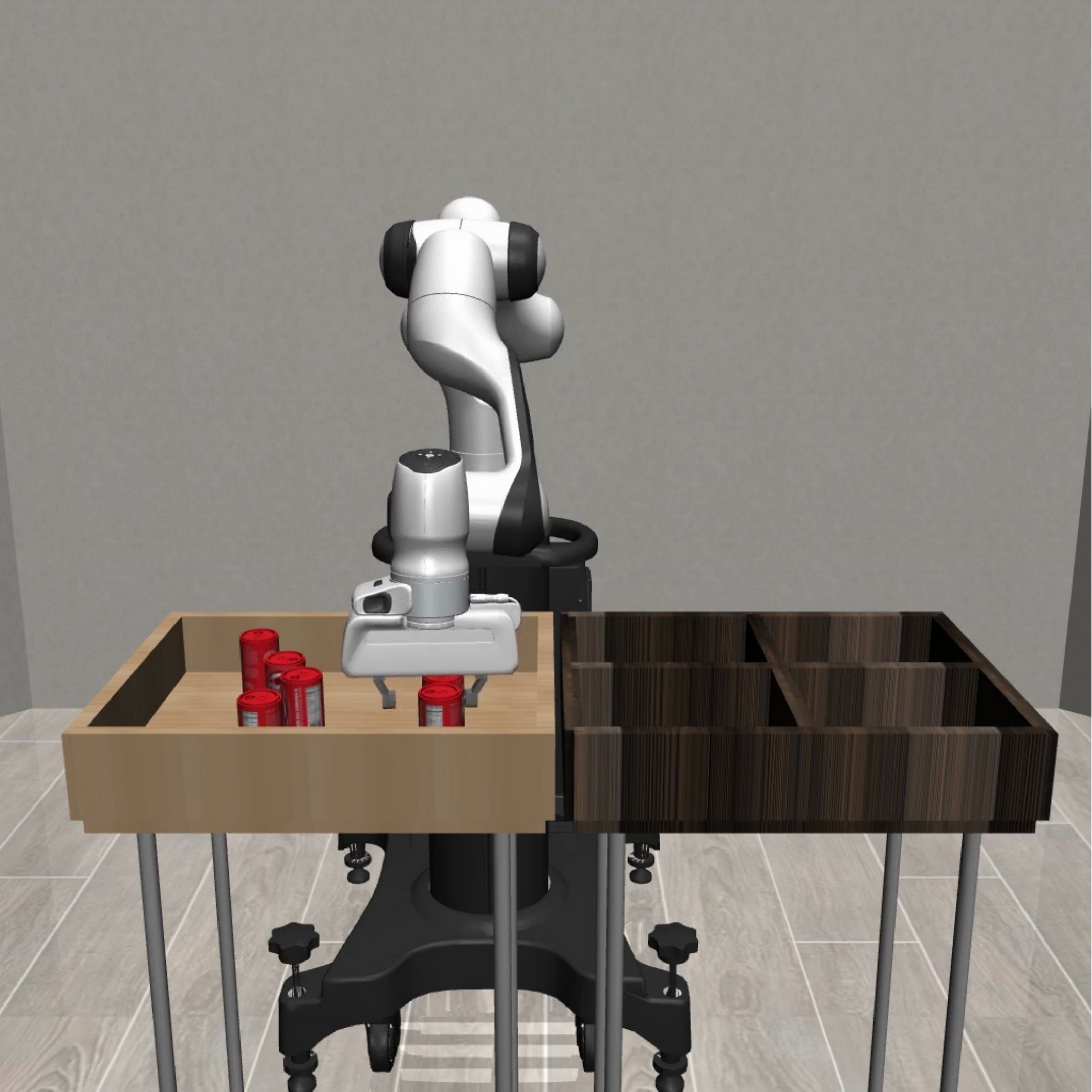}
    \end{subfigure}
    \hfil
    \begin{subfigure}{0.24\textwidth}
        \centering
        \includegraphics[width=\textwidth]{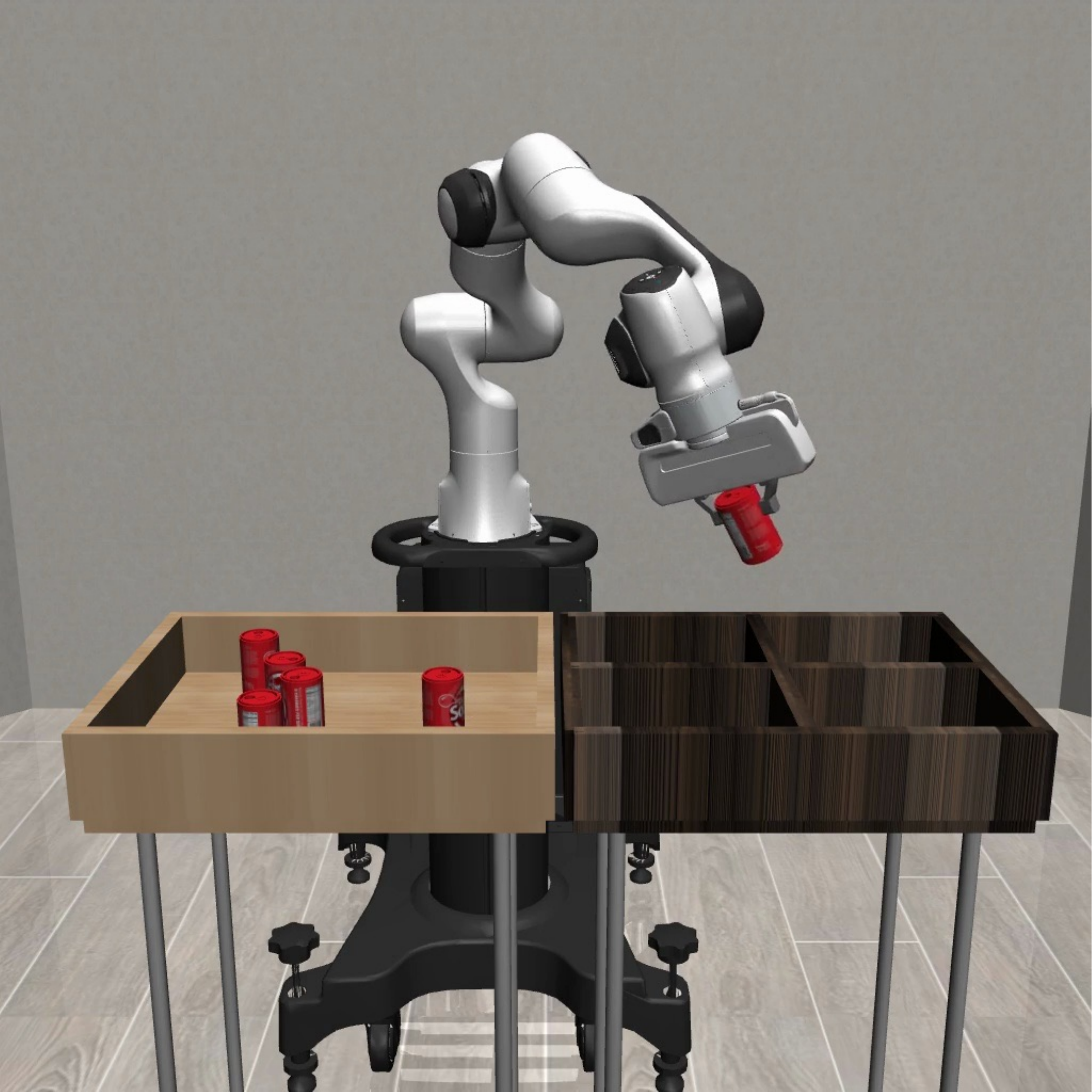}
    \end{subfigure}
    \hfil
    \begin{subfigure}{0.24\textwidth}
        \centering
        \includegraphics[width=\textwidth]{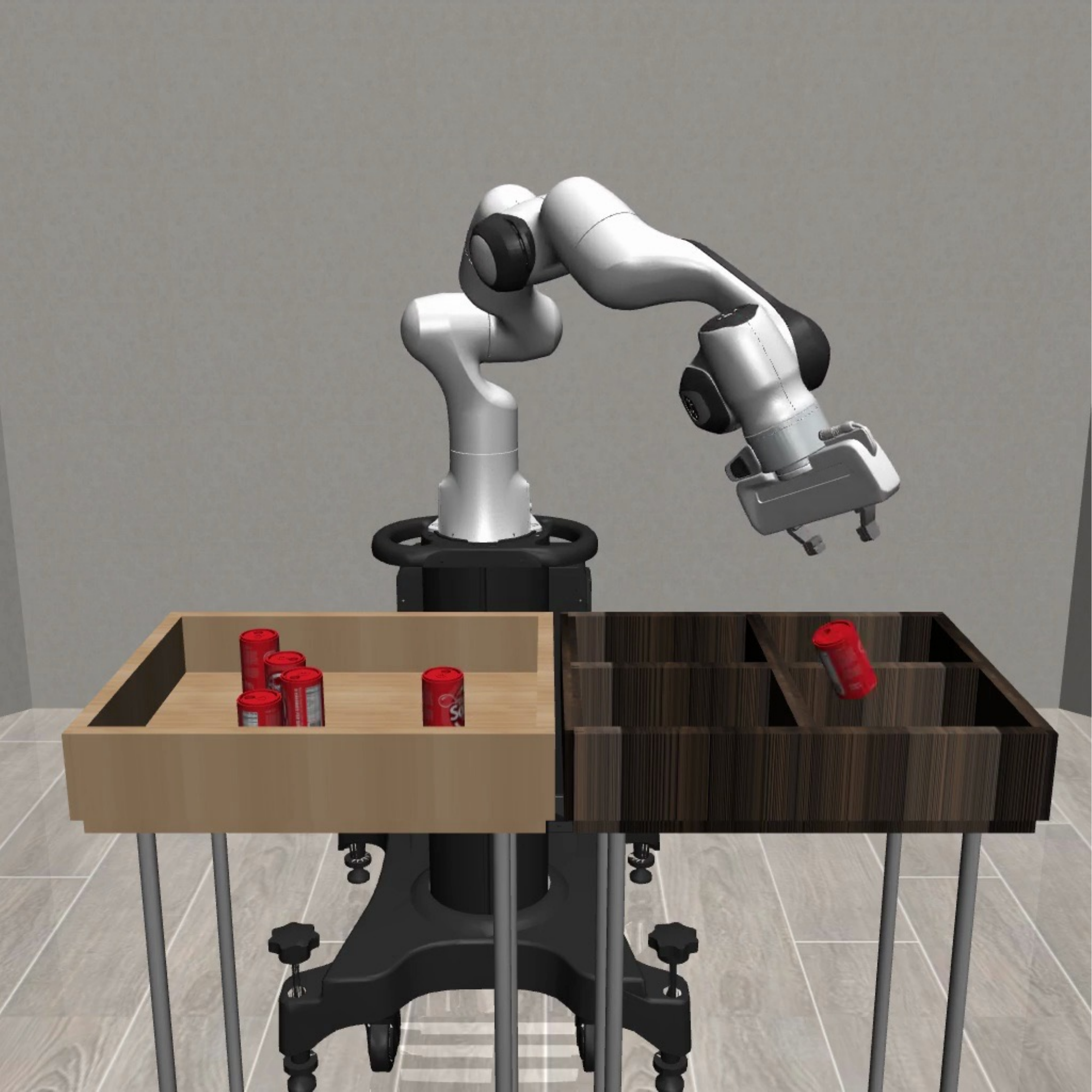}
    \end{subfigure}
    
    \caption{The pick-and-place task for preference labeling from the \textit{frontview}. The robot picks up a can from the left table and places it on the target table with four bins.}
    \Description{The pick-and-place task for preference labeling.}
    \label{fig:task}
\end{figure*}
As shown in \Cref{fig:task}, six cans are randomized on the left table, and the robot's goal is to pick up one of the cans and place it into the bins on the right.
A robot trajectory consists of states and actions, where the state is a vector of observations at each step (\ie each frame in the video), including the positions and velocities of all joints of the robot arm and all the objects collected using the robosuite framework \cite{zhuRobosuiteModularSimulation2020}.
The states are recorded in each frame, and the trajectory videos are recorded at $20$ frames per second.
In the basic preference labeling system, we utilized the K-means \cite{hartiganAlgorithm136KMeans1979} to perform cluster analysis on the states.
This analytical approach yielded $9$ representative samples from each cluster,
and the system randomly prompts the $36$ trajectory pairs during the study session.

\subsubsection{Procedure}
After getting familiar with the pick-and-place task and the labeling system,
the participants labeled $36$ pairs of robot pick-and-place trajectories through the system with their screen recorded.
The total labeling time ranges from $6$ to $22$ minutes ($M = 15.4$, $SD = 4.0$). 
After the labeling session, we conducted a retrospective think-aloud study \cite{ellingRetrospectiveThinkaloudMethod2011} to let the participants recall their mental process using the screen recording,
so that we can understand the way that they compare the trajectories.
Finally, we conducted semi-structured interviews with the participants to gain insights into: 
(a) the challenges they encounter when using the trajectory comparison system, 
(b) their criteria and features for comparing robot trajectories, 
(c) potential additional challenges they may face in scenarios involving labeling more pairs, and 
(d) their requirements for the trajectory preference labeling system.
The linked document \footnote{\url{https://bit.ly/3Hzil21}} presents the details of the formative study including demographic questions,  procedures, and semi-structured interview questions.

All the study sessions were fully recorded and transcribed for thematic analysis.
Two researchers independently coded the transcripts and got $22$ initial codes under $3$ themes, and we discussed the coding results in our research team meetings to categorize or divide the initial codes into $9$ codes and $19$ subcodes and reach a consensus.
Three themes were identified from the thematic analysis:
(1) what are the criteria and corresponding features that labelers care about in comparing robot trajectories,
(2) challenges in the preference elicitation process, and
(3) needs of the trajectory preference labeling system.
The formative study's findings are summarized in the following subsections.

%% file: sections/formative-study/criteria-features.tex
\subsection{Criteria and Features}
\label{sec:formative-study-criteria-features}
Participants compared two robot task trajectories based on observed criteria and features in the trajectory videos, which sheds light on the first part of RQ1: how labelers compare robot trajectories.
While prioritizing these criteria and features is challenging (described in \Cref{sec:formative-study-c1}), most participants considered the criteria and corresponding features in a hierarchical order.
During our interviews, we asked participants to suggest additional criteria and features they would consider for comparing trajectories other than those mentioned in the retrospective think-aloud.
After that, we presented a list of criteria and features summarized from other literature, such as related human values in \cite{yuanSituBidirectionalHumanrobot2022a} and evaluation metrics from \cite{dobisEvaluationCriteriaTrajectories2022}.

Some participants (P01, P03, P05, P12) admitted they would not consider certain specialized robot arm metrics from \cite{dobisEvaluationCriteriaTrajectories2022}, finding them hard to interpret or not important from their perspective.
For example, P01 ignored the arm and the tilting angle and cared more about whether the object was safe or the task was completed within an acceptable time, \etc. 
Similarly, P12 expressed his perspective, stating that judging the system's industrial performance might not be straightforward and could be subjective based on their intuitive point of view.
The way that participants compare the trajectories can reflect the mismatch between the preferences of common users (from high-level criteria) and the evaluation metrics in the robotic specializations (from low-level metrics).
Thus, rather than using features obtained directly from metrics in the previous literature, we summarized a set of criteria and corresponding features mentioned by labelers in our thematic analysis in \Cref{tab:criteria-features}.
\begin{table*}
    \centering
    \caption{Criteria and corresponding features that labelers consider in comparing robot trajectories}
    
    \begin{tabular}{m{.11\textwidth}m{.2\textwidth}m{.56\textwidth}}
        \toprule
        \textbf{Criterion}          & \textbf{Feature} & {\textbf{Description}} \\
        \midrule
        \multirow{4}{*}{Safety}     &
        Collision                   &
        Contact(s) of cared objects.                                                                                 \\
        \cmidrule{2-3}
                                    &
        Distance                    &
        The distance(s) between the cared point(s) / line(s) / surface(s).                                           \\
        \cmidrule{2-3}
                                    &
        Contact force               &
        The contact force(s) between the cared objects.                                                              \\
        \midrule
        \multirow{5}{*}{Efficiency} &
        Speed                       &
        The speed(s) of the cared object(s).                                                                         \\
        \cmidrule{2-3}
                                    &
        Path Length                 &
        The length(s) of the path(s) of the cared object(s).                                                         \\
        \cmidrule{2-3}
                                    &
        Time                        &
        The total time and time used for cared path(s).                                                              \\
        \cmidrule{2-3}
                                    &
        Power Usage                 &
        The total power consumption.                                                                                 \\
        \midrule
        \multirow{5}{*}{Task Quality}
                                    &
        Speed Smoothness            &
        The sum(s) of absolute changes in the speed(s) of the cared object(s).                                       \\
        \cmidrule{2-3}
                                    &
        Trajectory Smoothness       &
        The smoothness score(s) of the path(s) of the cared object(s).                                               \\
        \cmidrule{2-3}
                                    &
        Orientation                 &
        The relative orientation(s) between the gripper(s) and the target(s).                                        \\
        \cmidrule{2-3}
                                    &
        Grasp   Position            &
        The relative position(s) between the gripper(s) and the grasped target(s).                                   \\
        \bottomrule
    \end{tabular}
    \label{tab:criteria-features}
\end{table*}

These features are summarized using generalizable terms that can be applied to other robot tasks. For example, in this pick-and-place task, the \textit{distance} feature can be measured by the highest distance between the can and the table and the nearest distance between the can and the table edge, \etc.
All participants mentioned either \textit{safety} or \textit{efficiency} as their top priority criteria.
The \textit{safety} criterion is related to \textit{collision}, \textit{distance}, and \textit{contact force} features.
The \textit{efficiency} criterion is related to \textit{speed}, \textit{path length}, \textit{time}, and \textit{power usage} features.
For \textit{power usage} and \textit{contact force}, they are not observable from the trajectory videos, but participants mentioned that they would consider them if provided with the information.
The participants recognize the task quality criterion less, and it is related to \textit{speed smoothness}, \textit{trajectory smoothness}, \textit{orientation}, and \textit{grasp position} features.
However, participants talked about several features related to the task quality criterion in their think-aloud, such as smoothness (P01, P03, P04, P05, P08, P12), stableness (P06, P07, P08), object orientation (P02, P05, P06, P11, P12), gripper orientation (P06, P09), or grasp approach (P02, P03, P08, P11, P12).
For example, P06 thought \textit{lying flat down would be better} when comparing the can orientation of two trajectories.

%% file: sections/formative-study/challenges.tex
\subsection{Challenges}
\label{sec:formative-study-challenges}

This subsection answers RQ1 about the challenges during the preference elicitation process.
We categorize participants' challenges when comparing robot trajectories into three aspects (\Cref{fig:C-DR-DF}'s ``Challenges'' column).

\subsubsection{{C1 - Difficulty in forming criteria}}
\label{sec:formative-study-c1}

Forming criteria for comparing robot trajectories is the first step of the preference elicitation process, which is challenging for participants from the following two aspects:
\textit{C1.1 - unclear coverage of criteria features} and
\textit{C1.2 - unclear value distribution of each feature}.
On the one hand, participants (P01-P07, P12) knew too little about the performance of this robot to build a set of criteria covering all the new situations.
For example, P05 first felt that \textit{one of the essential things was that the robot could not knock the can down, and the robot probably will not be able to grab it next time}. But then he found that the robot could push the can down and grab it, which made him change the priority of the feature \textit{collision}.
On the other hand, participants (P02, P05, P08, P09) found it difficult to estimate the value distribution of each feature. 
Thus, they were not sure about the importance of each feature.
For example, many participants found it difficult to estimate the value distribution of the feature \textit{path length} because they did not know the range of the path length of the gripper or the can, which affected the priorities of the criterion \textit{efficiency}.

\subsubsection{{C2 - Overlooking trajectory details}}
\label{sec:formative-study-c2}

Participants found it challenging to pay attention to the details of the trajectories.
The challenge includes two factors:
\textit{C2.1 - the lack of robotic understandings} and
\textit{C2.2 - the unobservability of video details}.
Due to the knowledge limitation, participants (P01, P02, P03, P06, P09, P12) found figuring out some robotic details challenging. 
For example, P03 admitted not having enough knowledge about the robotic arm to judge its power usage without knowing which joint is more power-consuming.
Besides, observing the non-salient details of the trajectories from the videos is also not accessible
and some features like \textit{contact force} are not observable from the trajectory videos (Mentioned by P01-P03, P05-P07). 
For example, during retrospective think-aloud, P07 did not notice a minor collision between the can and the table.

\subsubsection{{C3 - Difficulty in maintaining focus}}
\label{sec:formative-study-c3}

The labeling process is time-consuming, and many participants failed to maintain focus during the labeling process.
The following two factors contribute to this challenge:
\textit{C3.1 - mental fatigue} and
\textit{C3.2 - lack of feedback}.
Labeling similar trajectories showing the same simple task, participants (P01, P02, P04) felt bored and mentally exhausted.
For example, P04 said that \textit{``I'm going to be honest with you, this was incredibly tedious.''}
Additionally, participants (P03, P05, P07) anticipated knowing their progress, the quality of their labeling, and improvements in the robot's performance but did not receive any feedback, which hindered their ability to stay focused.
Many participants kept asking how many pairs were left during the labeling process.

%% file: sections/formative-study/design-requirements.tex
\subsection{Derived Design Requirements}
\label{sec:formative-study-requirements}


The last part of RQ1 is what labelers need for the trajectory preference labeling system.
We derive the design requirements and features of the trajectory preference labeling system from the coded challenges and needs.
The design requirements and features are in the ``Design Requirements'' and ``Design Features'' columns of \Cref{fig:C-DR-DF}, with the relations between columns shown in the connecting lines.

\subsubsection{{DR1. Dynamically arranging the labeling order to balance difficulty}}
\label{sec:formative-study-dr1}

{DR1} aims to address the challenges of {C1.1} and {C2.2}, which involve helping labelers establish a consistent criteria system and manage the difficulty of labeling.
Participants found it challenging to label similar trajectory pairs without establishing consistent criteria.
Therefore, they wanted to label trajectory pairs with distinct features to establish consistency at the beginning of the labeling task.
To address this challenge, {DR1} requires the system to prompt labelers to label trajectory pairs with distinct features at the initial stage, followed by trajectory pairs with similar features after labelers have viewed a more comprehensive range of feature varieties.
Furthermore, we propose design features \textit{DF1. Trajectory data clustering based on pre-defined feature vectors} and \textit{DF3. Dynamic trajectory pair prompting based on clustering results and labeling progress} to satisfy {DR1}.

\subsubsection{{DR2. Highlighting $~$ noteworthy $~$ trajectory $~$ features $~$ and $~$ keyframes}}
\label{sec:formative-study-dr2}

The goal of {DR2} is to address the challenges of {C1.2} and {C2}, \ie help labelers identify features and keyframes of trajectories.
More information about features not easily identified from the videos, such as \textit{power usage} and \textit{contact force}, can help participants prioritize the features more accurately.
Participants also wished for the ability to jump to specific keyframes to aid in recalling trajectory features and observing details.
To fulfill this need, {DR2} requires the system to provide more information, including distributions of non-salient features and keyframes, to assist labelers in making comparisons.
We propose design features \textit{DF2. Feature-based keyframe extraction from each trajectory} and \textit{DF5. Adaptive display of keyframes and feature outlier distributions} to satisfy {DR2}.

\subsubsection{{DR3. Real-time attention monitoring and feedback provision}}
\label{sec:formative-study-dr3}

{DR3} is to tackle the challenge {C3}, which is to help labelers focus on the labeling task and provide feedback to labelers.
Participants mentioned that they would like to know how much progress they have made and how well they are doing.
Thus, we use participants' consistency to proxy their attention and fatigue to monitor their performance and provide feedback.
To realize {DR3}, \textit{DF4. Consistency check and reminders} is proposed to monitor labelers' attention and provide feedback to participants.

%% file: sections/preference-collection-system.tex
\section{\tool: System Design and Implementation}
\label{sec:system}

This section provides a detailed illustration of the design and implementation of our \tool{} system, 
which comprises three main subsections: the dataset, the server component, and the user interface component (\Cref{fig:C-DR-DF}'s ``Components'' columns).
These components are implemented to fulfill design features as shown in lines connected to ``Design Features'' in \Cref{fig:C-DR-DF}. 

\input{sections/implementation/dataset.tex}
\input{sections/implementation/server.tex}
\input{sections/implementation/frontend.tex}

%% file: sections/implementation/dataset.tex
\subsection{Feature-Augmented Trajectory Dataset}
\label{sec:dataset}

\input{sections/floats/tab-feature-keyframe-measure.tex}

\input{sections/floats/fig-keyframes.tex}

\subsubsection{Dataset generation}
\label{sec:dataset-generation}
The pick-and-place task is common in real life and thus a good starting point for generating a dataset for preference labeling.
We create the simulation environment using the robosuite framework \cite{zhuRobosuiteModularSimulation2020}.
The \textit{Can} task of robomimic \cite{robomimic2021} dataset \footnote{\url{https://robomimic.github.io/docs/v0.2/datasets/robomimic_v0.1.html\#info}} contains $200$ successful proficient human demonstration trajectories (ph), $300$ successful multi-human demonstration trajectories (mh), $3900$ machine-generated trajectories (mg) and $100$ successful human demonstration trajectories paired with $100$ unsuccessful human demonstration trajectories (paired).
As the robomimic \textit{Can} task contains only a single can, we randomly simulated the initial positions of the other five cans on the left table as a new environment.
First, we use the actions from the robomimic dataset to generate successful trajectories for the robot to pick up and place one of the cans in our environment.
Then, we select trajectories where the robot's end-effector remained confined within the horizontal spatial constraints of the edges of two tables.
Furthermore, we remove human demonstration trajectories longer than $8$ seconds, which is the longest duration in machine-generated trajectories.
Finally, we get our dataset \dataset{} with $636$ successful trajectories, including $200$ ph, $117$ mh, $99$ paired and $220$ mg.
All participants in the formative study expressed a preference for the \textit{frontview}, \ie the camera view from the front, over other views.
Therefore, we continued to use the \textit{frontview} videos to demonstrate the trajectories.

\subsubsection{Feature and keyframe definitions}
\label{sec:feature-keyframe}
According to the general criteria and features that labelers care about in \Cref{sec:formative-study-criteria-features},
we define each feature and feature-based keyframes to be extracted for each trajectory in \dataset{} in \Cref{tab:feature-keyframe-measure}.
The detailed formulas for each feature in either time series or scalar values are in \Cref{tab:feature-keyframe-formula} of \Cref{sec:feature-formula-definitions}.
We stack the time series features together to form the criterion vector series, so that we can represent each criterion for the trajectory and calculate the similarity between trajectories for clustering.
We calculate the scalar values and extract keyframes to display the feature distribution on the interface, which is more intuitive for the labelers to compare the features of different trajectories.

\subsubsection{Criteria-based clustering}
\label{sec:clustering}

We create a criterion vector series for each criterion with the time series of corresponding features stacked together.
$\mathbf{safety}_i(t)$, $\mathbf{efficiency}_i(t)$, and $\mathbf{task\_quality}_i(t)$ denote the criterion vector series of three criteria, \textit{Safety}, \textit{Efficiency}, and \textit{Task Quality}, for trajectory $i$ at $t$-th step (\Cref{sec:feature-formula-definitions}).
We compute Dynamic Time Warping (DTW) distance matrices \cite{salvadorFastDTWAccurateDynamic2007} $\mathbf{D}_{safety}$, $\mathbf{D}_{efficiency}$, and $\mathbf{D}_{task\_quality}$. Each entry $\mathbf{D}_{safety}(i,j)$ denotes the DTW distance between the safety vector series
of trajectory $i$ and $j$.
$\mathbf{D}_{efficiency}(i,j)$ and $\mathbf{D}_{task\_quality}(i,j)$ are defined similarly.
Then we obtain the pairwise distance matrix $D$ using the weighted sum of the DTW distance matrices for each criterion to represent the similarity between trajectories:
$$
    \mathbf{D} = \frac{w_{s}\mathbf{D}_{safety} + w_{e}\mathbf{D}_{efficiency} + w_{t}\mathbf{D}_{task\_quality}}{w_{s} + w_{e} + w_{t}},
$$
We set equal weights for all criteria in this work.
In practice, designers can add weights to prioritize the criteria in the similarity calculation.
Then we obtain our feature-augmented, clustered dataset \dataset.

%% file: sections/floats/tab-feature-keyframe-measure.tex
\begin{table*}
    \centering
    \caption{Definitions for each feature and corresponding keyframes.}
    \setlength{\colonewidth}{\dimexpr .185\textwidth - 2\tabcolsep}
    \setlength{\coltwowidth}{\dimexpr .545\textwidth - 2\tabcolsep}
    \setlength{\colthreewidth}{\dimexpr .27\textwidth - 2\tabcolsep}
    \begin{tabular}{m{\colonewidth}m{\coltwowidth}m{\colthreewidth}}
        \toprule
        \textbf{Feature}
         &
        \textbf{Definition}
         &
        \textbf{Keyframe}
        \\
        \midrule
        \multicolumn{3}{c}{Safety}
        \\

        \midrule
        Collision
         &
        The number of contacts between all the objects, except those between robot fingers and the target can.
         &
        The frames with the collisions (\Cref{fig:collision1,fig:collision2})
        \\
        \cmidrule{2-3}
        Distance
         &
        \begin{tabular}[c]{@{}p{\coltwowidth}@{}}
            1) The minimum distances between the target can and each table edge. \\
            2) The maximum height of the target can from the table surface.
        \end{tabular}
         &
        \begin{tabular}[c]{@{}p{\colthreewidth}@{}}
            The frames of the can:                                                              \\
            1) with the minimum distance to all table edges (\Cref{fig:nearest_point_to_edge}); \\
            2) with the maximum height to the table (\Cref{fig:highest_point})
        \end{tabular}
        \\
        \cmidrule{2-3}
        Contact   force
         &
        The maximum force exerted by the robot end-effector.
         &
        N.A.
        \\
        \midrule
        \multicolumn{3}{c}{Efficiency}
        \\
        \midrule
        Speed
         &
        The average speed of the robot end-effector during the task.
         &
        N.A.
        \\
        \cmidrule{2-3}
        Path   Length
         &
        \begin{tabular}[c]{@{}p{\coltwowidth}@{}}
            1) The path length of the robot end-effector reaching the target can. \\
            2) The path length of the end-effector grasping the target can.       \\
            3) The path length of the can from pick-up to placement.
        \end{tabular}
         &
        N.A.
        \\
        \cmidrule{2-3}
        Time
         &
        \begin{tabular}[c]{@{}p{\coltwowidth}@{}}
            1) The time of the robot end-effector reaching the target can. \\
            2) The time of the end-effector grasping the target can.       \\
            3) The time from pick-up to placement.                         \\
            4) The total time of the manipulation task.
        \end{tabular}
         &
        \begin{tabular}[c]{@{}p{\colthreewidth}@{}}
            The frames of:                                   \\
            1) the pick-up point (\Cref{fig:pick_up_point}); \\
            2) the release point (\Cref{fig:release_point}).
        \end{tabular}
        \\
        \cmidrule{2-3}
        Power Usage
         &
        \begin{tabular}[c]{@{}p{\coltwowidth}@{}}
            The sum of the absolute values of the joint rotations as a proxy for power usage \cite{dobisEvaluationCriteriaTrajectories2022}, \ie \\
            $\quad pseudo\_cost =   \sum_{i=1}^{n} |q_i|,$                                                                                       \\
            where $q_i$ is the joint rotation of the $i$-th joint, and $n$ is the number of joints.
        \end{tabular}
         &
        N.A.
        \\
        \midrule
        \multicolumn{3}{c}{Task Quality}
        \\
        \midrule
        Speed Smoothness
         &
        \begin{tabular}[c]{@{}p{\coltwowidth}@{}}
            The sum of the absolute values of the end-effector's acceleration, \ie \\
            $\quad speed\_smoothness =   \sum_{j=1}^{s} \sqrt{\mathbf{a_j}^2},$    \\
            where $\mathbf{a_j}$ is the $6$-dimension acceleration vector of the end-effector at the $j$-th state and $s$ is the number of states in the trajectory.
        \end{tabular}
         &
        N.A.
        \\
        \cmidrule{2-3}
        Trajectory Smoothness
         &
        \begin{tabular}[c]{@{}p{\coltwowidth}@{}}
            The sum of angles between the displacements between adjacent states, \ie
            \\
            $\quad trajectory\_smoothness =   \sum_{j=1}^{s} \arccos{\frac{\mathbf{x_{j}} \cdot   \mathbf{x_{j+1}}}{|\mathbf{x_{j}}| |\mathbf{x_{j+1}}|}},$ \\
            where $\mathbf{x_{j}}$ is the $3$-dimension end-effector displacement vector between the $(j-1)$-th state and the $j$-th state.
        \end{tabular}
         &
        N.A.
        \\
        \cmidrule{2-3}
        Orientation
         &
        The maximum relative angle between the can's and the end-effector's orientation.
         &
        N.A.
        \\
        \cmidrule{2-3}
        Grasp   Position
         &
        The relative position vector with the largest distance between the can's and the end-effector's center during the grasping time.
         &
        N.A.
        \\
        \bottomrule
    \end{tabular}
    \label{tab:feature-keyframe-measure}
\end{table*}

%% file: sections/floats/fig-keyframes.tex
\begin{figure*}
    \centering
    \begin{subfigure}{0.3\textwidth}
        \centering
        \includegraphics[width=\textwidth]{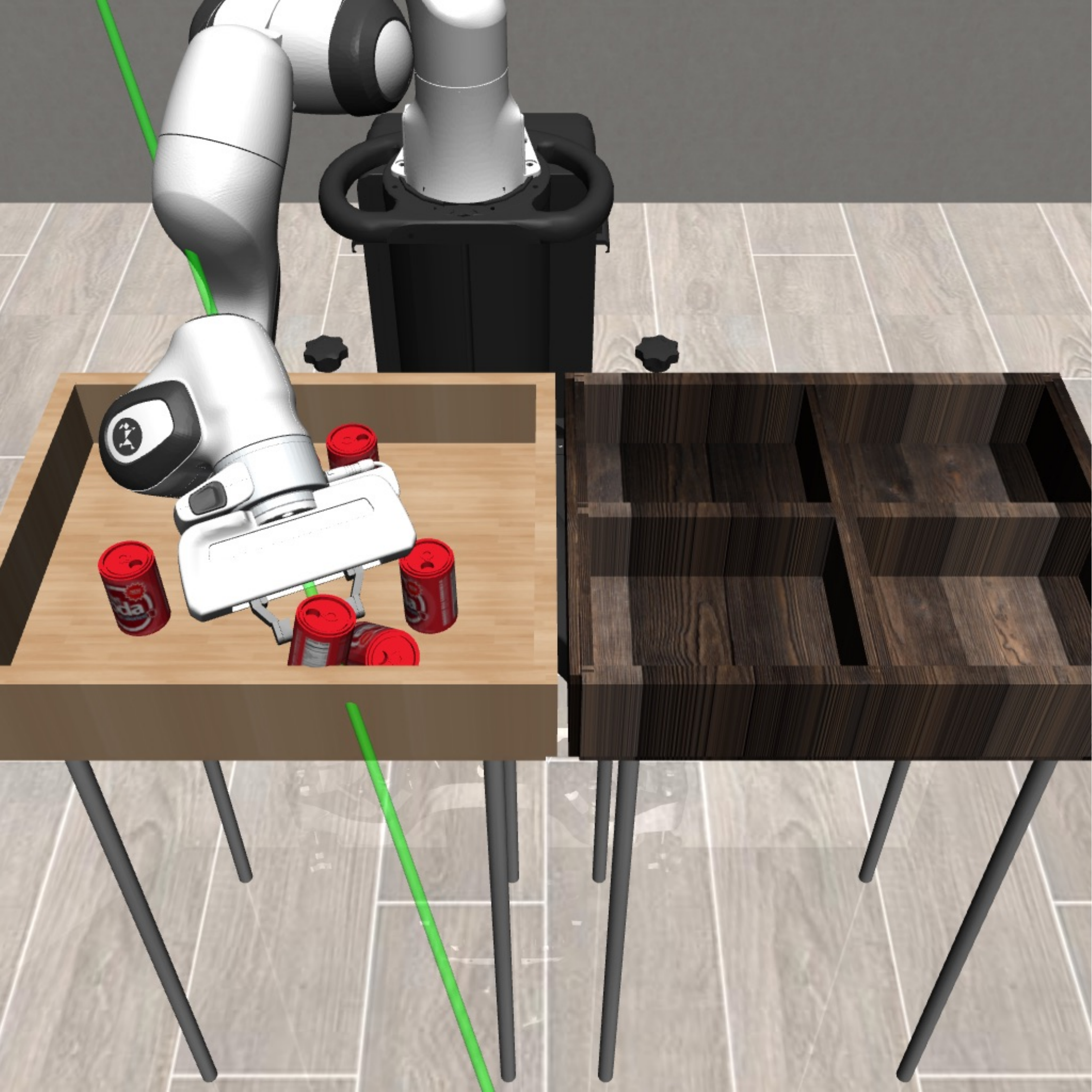}
        \subcaption{Collision with other cans (\textit{agentview})}
        \label{fig:collision1}
    \end{subfigure}
    \hfil
    \begin{subfigure}{0.3\textwidth}
        \centering
        \includegraphics[width=\textwidth]{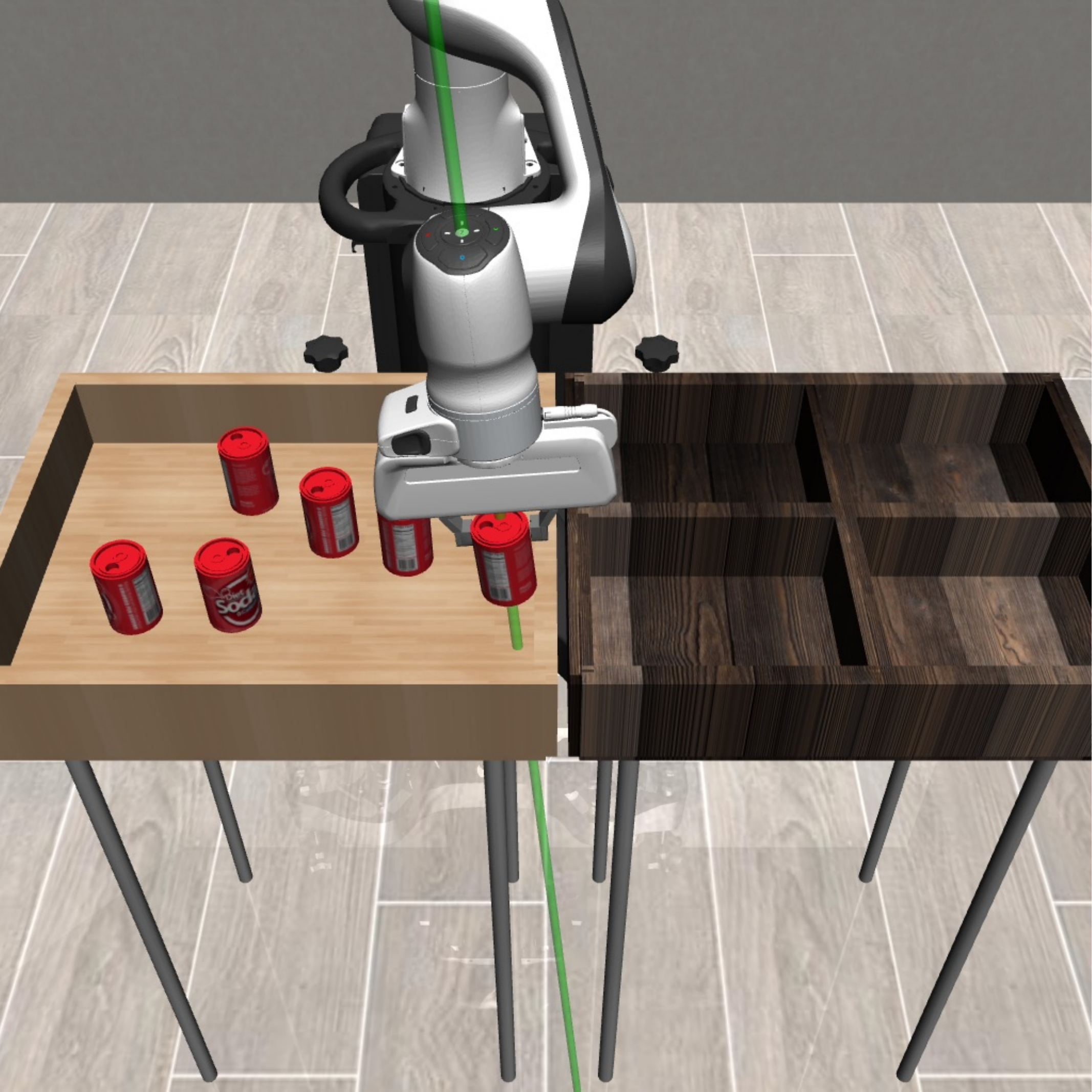}
        \subcaption{Collision with the table (\textit{agentview})}
        \label{fig:collision2}
    \end{subfigure}
    \hfil
    \begin{subfigure}{0.3\textwidth}
        \centering
        \includegraphics[width=\textwidth]{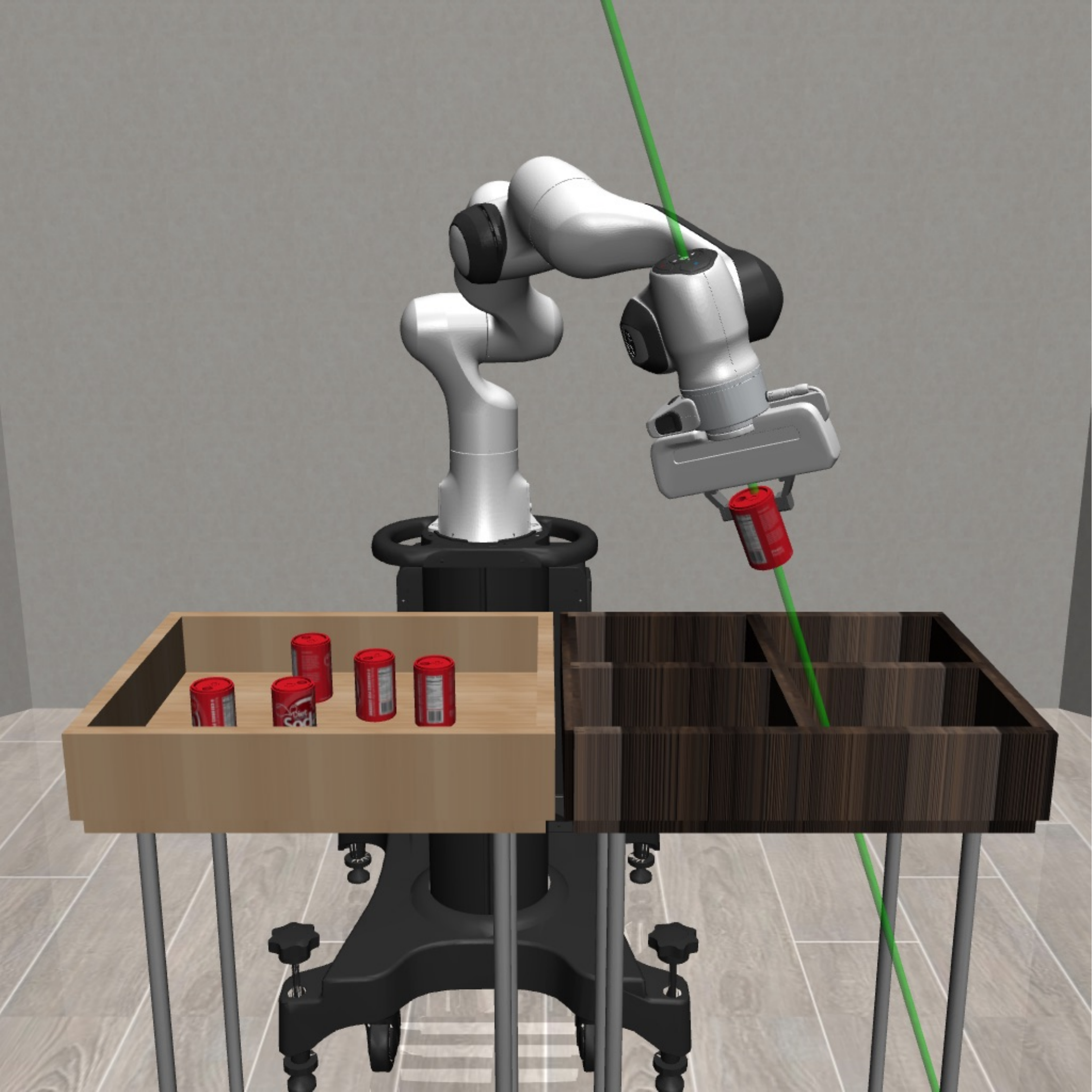}
        \subcaption{Highest point to table (\textit{frontview})}
        \label{fig:highest_point}
    \end{subfigure}
    \medbreak
    \begin{subfigure}{0.3\textwidth}
        \centering
        \includegraphics[width=\textwidth]{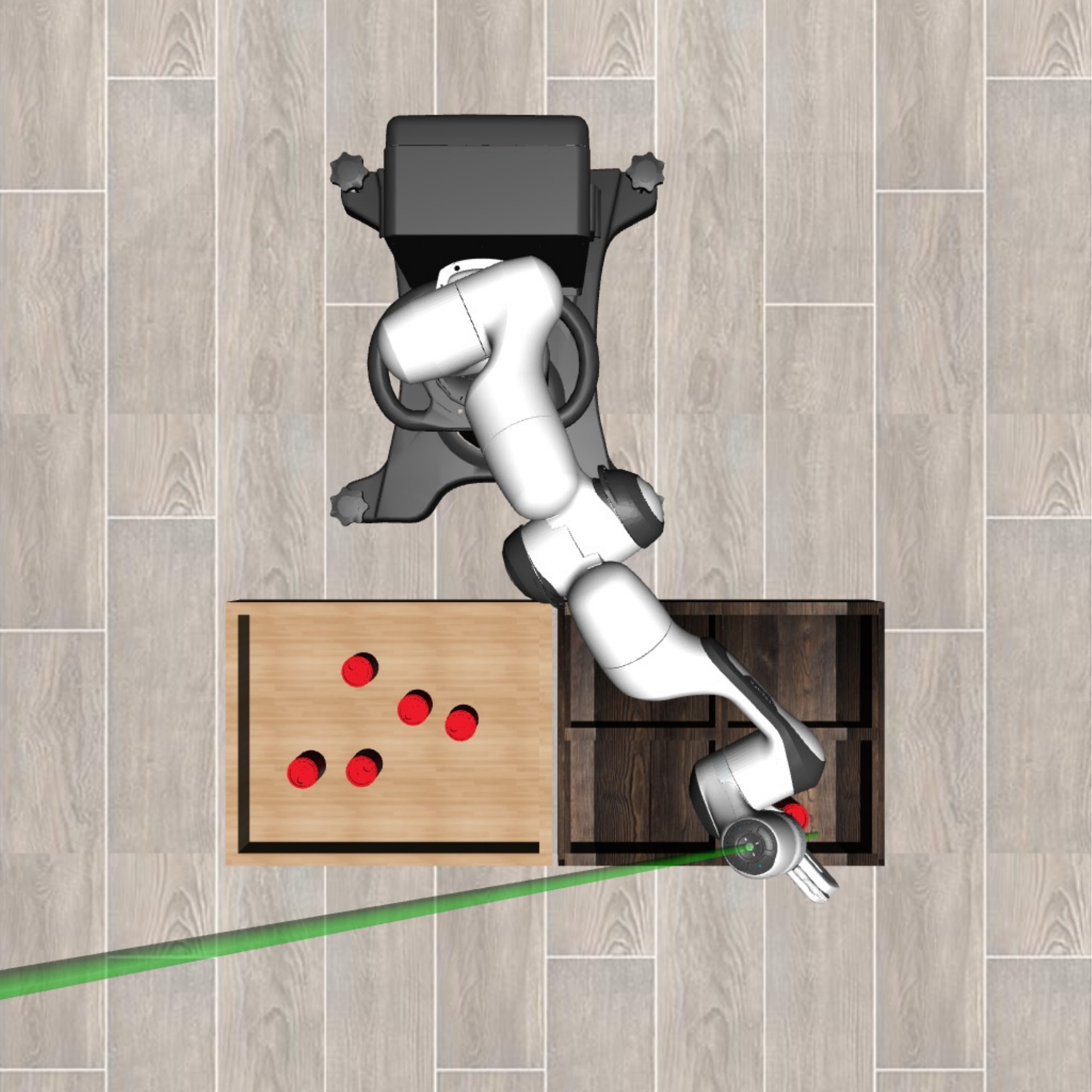}
        \subcaption{Nearest point to all edges (\textit{birdview})}
        \label{fig:nearest_point_to_edge}
    \end{subfigure}
    \hfil
    \begin{subfigure}{0.3\textwidth}
        \centering
        \includegraphics[width=\textwidth]{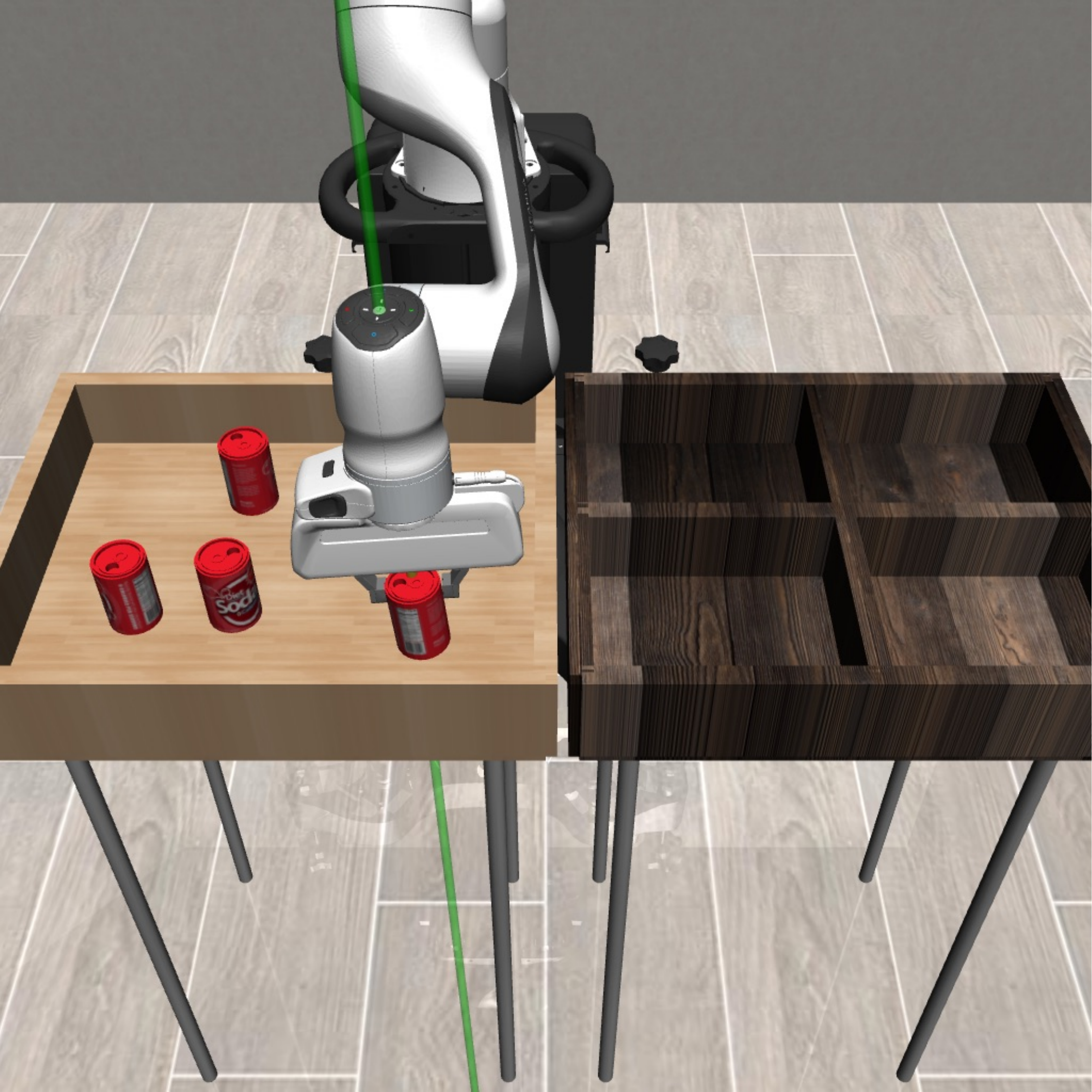}
        \subcaption{Pick up point (\textit{agentview})}
        \label{fig:pick_up_point}
    \end{subfigure}
    \hfil
    \begin{subfigure}{0.3\textwidth}
        \centering
        \includegraphics[width=\textwidth]{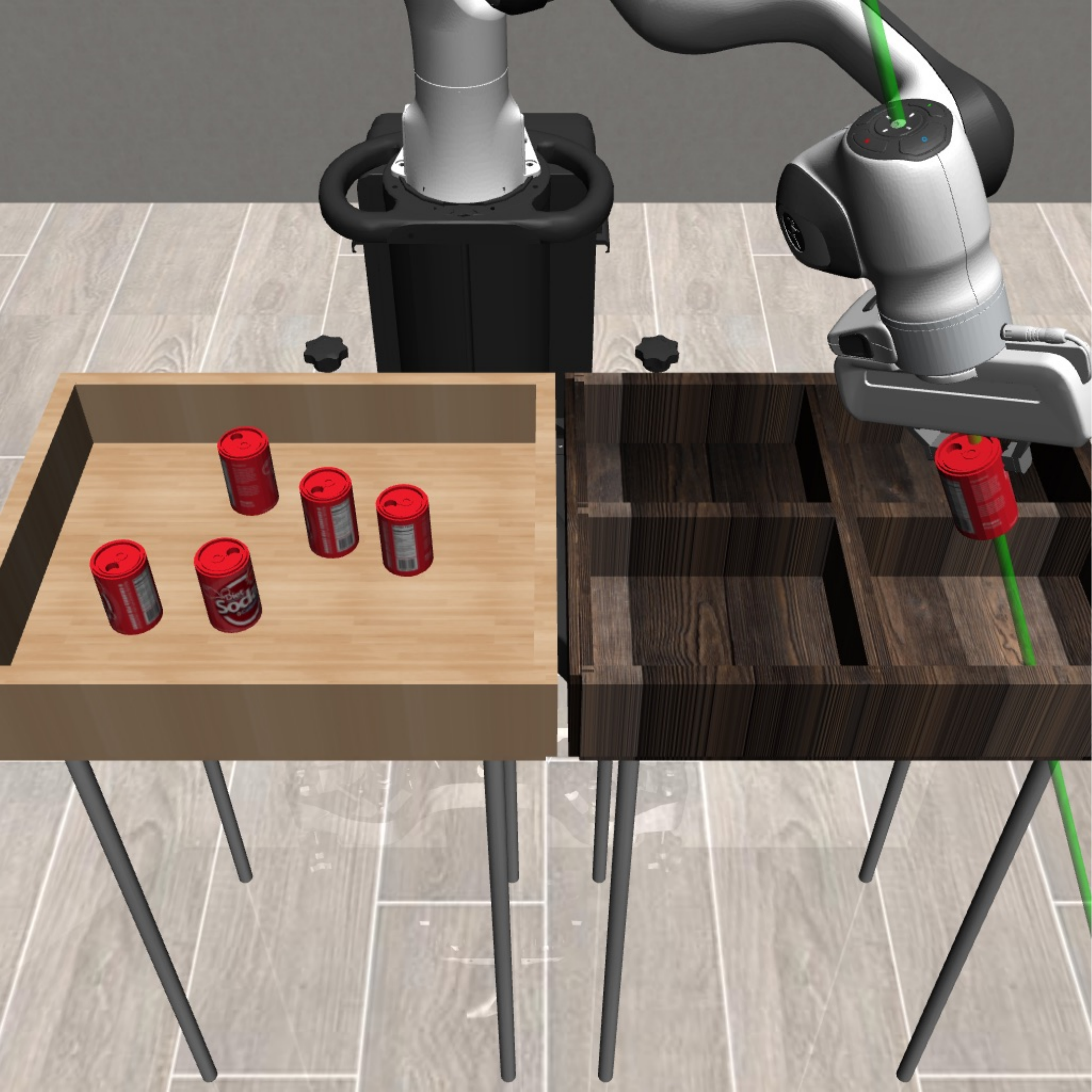}
        \subcaption{Release point (\textit{agentview})}
        \label{fig:release_point}
    \end{subfigure}
    
    \caption{{Examples of five kinds of keyframes, each captured from the best view to observe the corresponding feature.}}
    \Description{Examples of keyframes.}
    \label{fig:keyframes}
\end{figure*}

%% file: sections/implementation/server.tex
\subsection{Server Component}
\label{sec:implementation-server}

\input{sections/floats/tab-ranking-metrics.tex}

\subsubsection{Prompting strategy}
\label{sec:implementation-server-prompting}
We use the metrics in \Cref{tab:ranking-metrics} to calculate a dynamic ranking score of all trajectory pairs for each labeler.
These metrics are dynamically transformed to rank-based scores \cite{wolfe2009rank} to be averaged to the final ranking score.
The larger the ranking score, the higher the priority of the trajectory pair to be prompted to the user.

At the initial stage, when the users have not seen all the clusters,
the server chooses all trajectory pairs with none of their clusters covered
and ranks them based on the mean of other metrics to prompt the user.
This initial stage aims to help the user to have a whole glance at the features from different clusters, reducing the difficulty in forming criteria (\textit{C1}).
After the user has seen all the clusters, the server will choose the trajectory pairs with the least label skewness score and rank them based on other metrics to prompt the user.
These metrics balance the labeling difficulty, user familiarity and disagreements (\textit{DR1}).

\subsubsection{Consistency checking}
\label{sec:implementation-server-consistency-checking}
We arrange a consistency-checking round every $10$ normal prompting rounds, except that the first consistency-checking round is arranged after prompting $15$ unique trajectory pairs.
For each consistency-checking round, we randomly select one labeled trajectory and prompt it to the labeler to check the consistency.
If the labeler's preference label is consistent with the previous preference, we will prompt an encouraging message to the labeler:
\begin{quote}
    According to our record so far, you have been rather careful and thorough in the past labeling sessions!
    Good job! Take a break if needed and keep on the good work.
\end{quote}
Otherwise, we will prompt a message to remind the labeler to take a rest and be more careful:
\begin{quote}
    Feeling tired? Take a break if necessary and please stay attentive in the following sessions.
\end{quote}
This consistency-checking mechanism follows \textit{DF4. Consistency check and reminders} aiming to fulfill \textit{DR3. Real-time attention monitoring and feedback provision} that deals with challenge \textit{C3 - Difficulty in maintaining focus}.

%% file: sections/floats/tab-ranking-metrics.tex
\begin{table*}
    \centering
    \caption{Metrics for dynamic ranking scores}
    
    \begin{tabular}{m{\dimexpr .23\textwidth - 2\tabcolsep}m{\dimexpr .62\textwidth - 2\tabcolsep}m{\dimexpr .125\textwidth - 2\tabcolsep}}
        \toprule
        \textbf{Metric}
         &
        \textbf{Definition}
         &
        \textbf{Ranking}
        \\
        \midrule
        \textit{Cluster Coverage}
         &
        The portion of prompted trajectories from each cluster.
         &
        Descending
        \\
        \cmidrule{1-3}
        \textit{Combination Familiarity}
         &
        The portion of labeled trajectory pairs from each cluster combination.
         &
        Descending
        \\
        \cmidrule{1-3}
        \textit{Pair Similarity}
         &
        The distance between the normalized feature vectors (\Cref{sec:feature-keyframe}) from each trajectory pair.
         &
        Ascending
        \\
        \cmidrule{1-3}
        \textit{Pair Disagreement}
         &
        The variance \cite{christianoDeepReinforcementLearning2017, shinBenchmarksAlgorithmsOffline2022} of all users' preference scores ($preference\_score \in \{0, 0.5, 1\}$) for each trajectory pair.
         &
        Ascending
        \\
        \cmidrule{1-3}
        \textit{Cluster Disagreement}
         &
        The variance of all users' preference scores ($preference\_score \in \{0, 0.5, 1\}$) for all trajectory pairs in each cluster combination.
         &
        Ascending
        \\
        \cmidrule{1-3}
        \textit{Label Skewness}
         &
        The number of users that labeled each trajectory pair.
         &
        Descending
        \\
        \bottomrule
    \end{tabular}
    \label{tab:ranking-metrics}
\end{table*}

%% file: sections/implementation/frontend.tex
\subsection{User Interface Component}
\label{sec:implementation-ui}

\input{sections/floats/fig-ui-overview.tex}

\begin{figure}
    \centering
    \includegraphics[width=.3\textwidth]{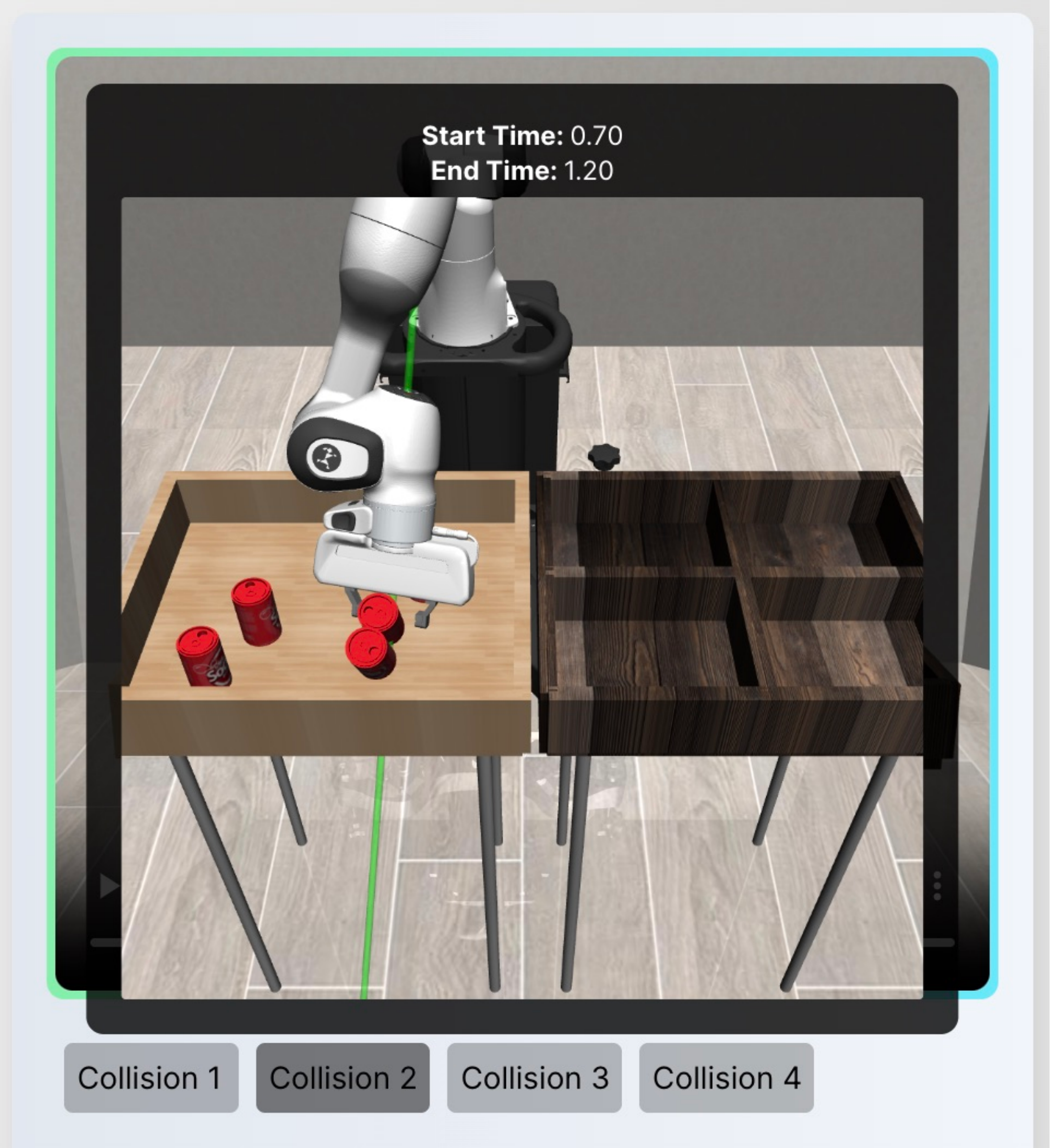}
    \caption{{Collision keyframe preview when hovering on the buttons.}}
    \label{fig:ui-collision-hover}
\end{figure}

\begin{figure}
    \centering
    \includegraphics[width=.475\textwidth]{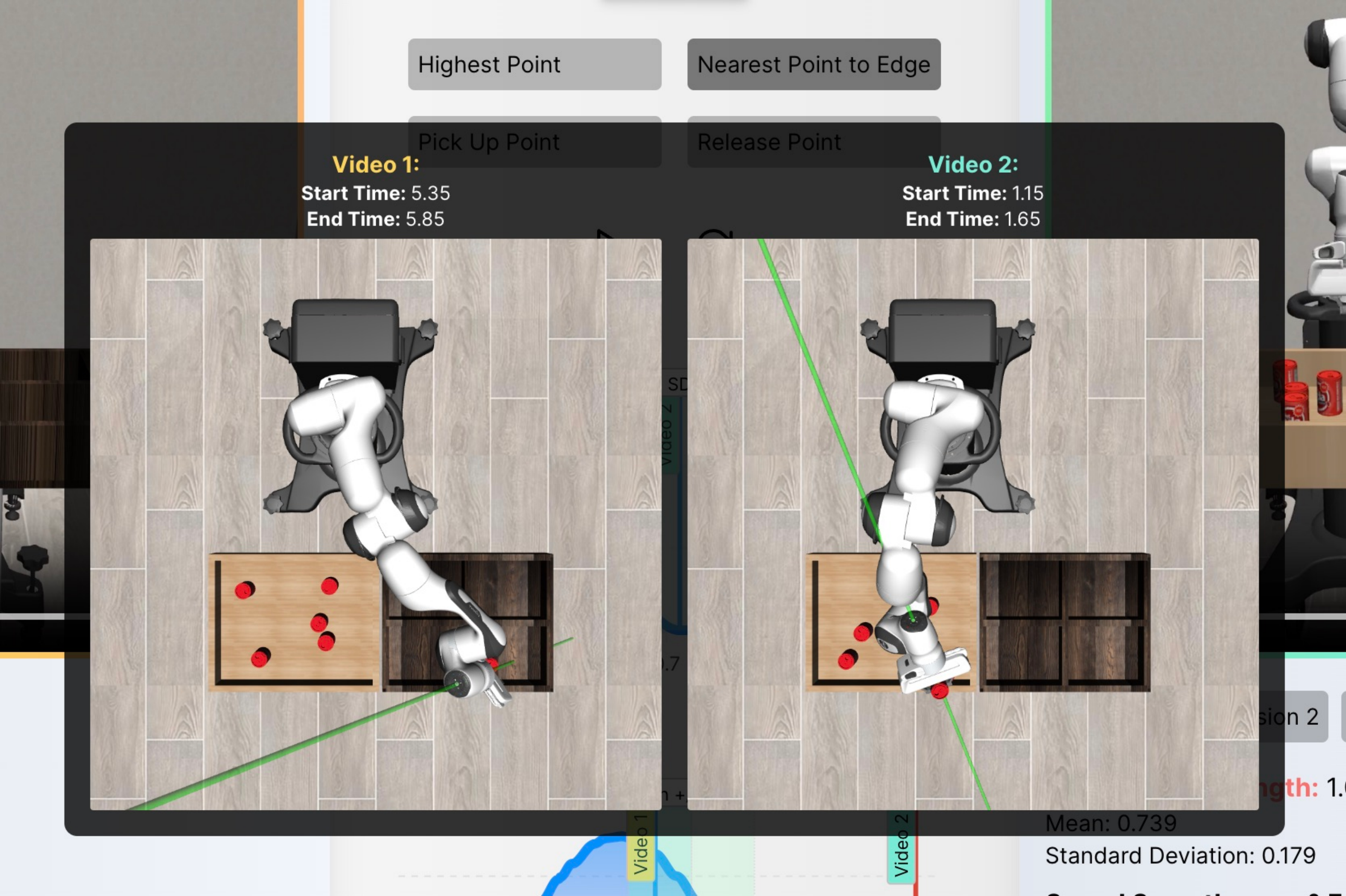}
    \caption{{Common keyframe preview when hovering on the buttons.}}
    \label{fig:ui-keyframe-hover}
\end{figure}

\begin{figure}
    \centering
    \includegraphics[width=.4\textwidth]{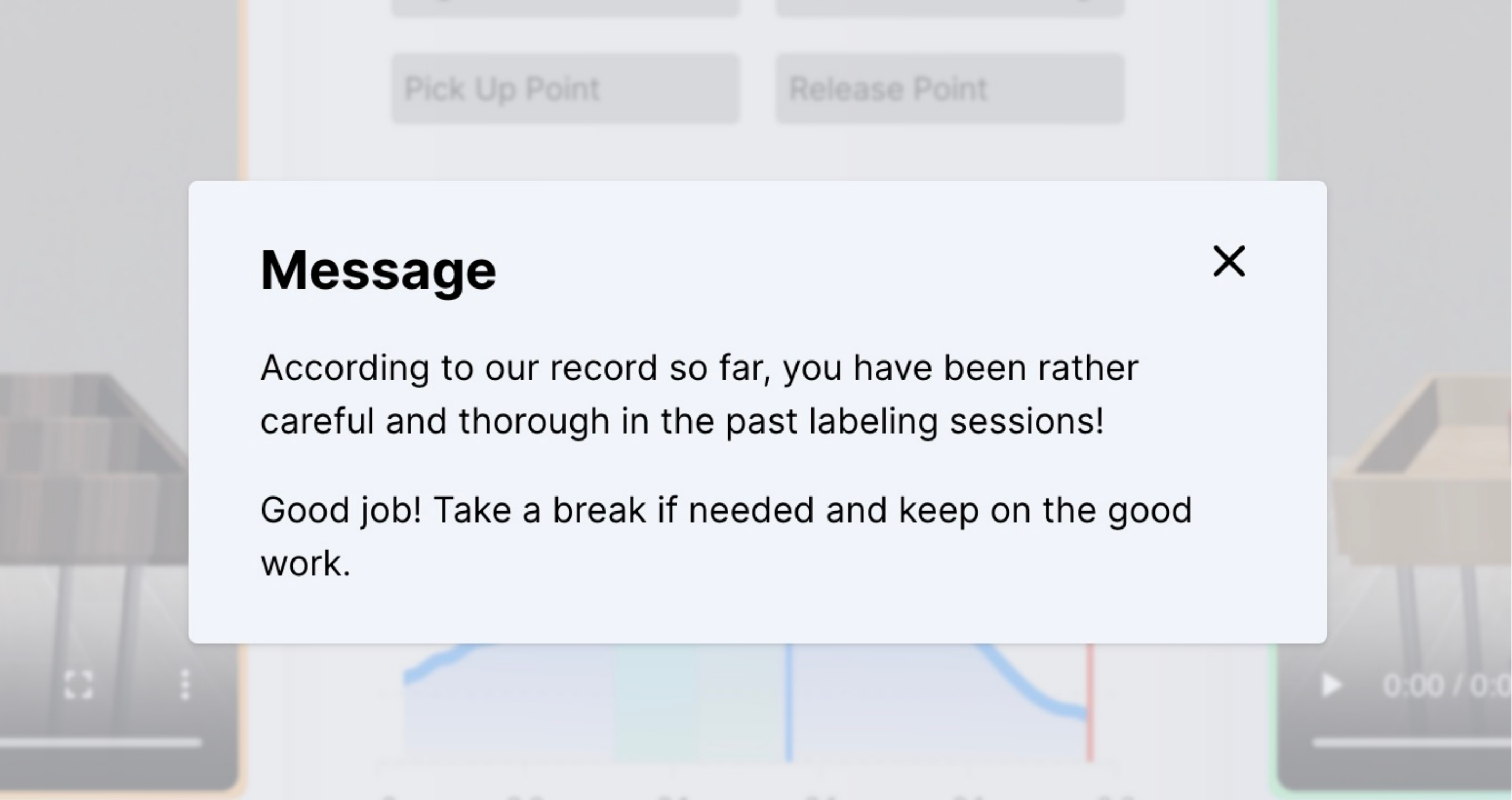}
    \caption{{Encouraging messages.}}
    \label{fig:ui-encouraging-message}
\end{figure}

\begin{figure}
    \centering
    \includegraphics[width=.4\textwidth]{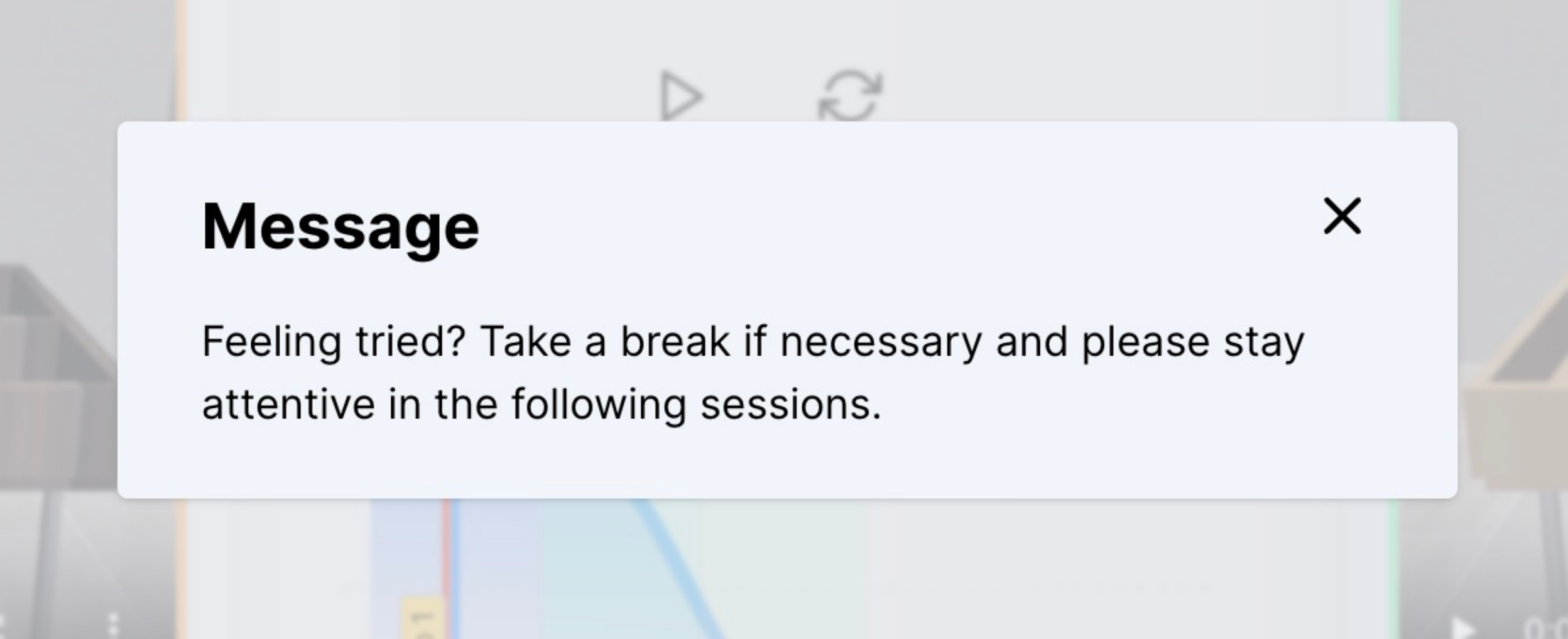}
    \caption{{Rest reminder messages.}}
    \label{fig:ui-taking-rest-message}
\end{figure}

We design the interface of \tool{} based on the interface of \citet{christianoDeepReinforcementLearning2017} and add our design features related to the user interface component (\textit{DF4} and \textit{DF5})
The user interface overview and descriptions for all components are in \Cref{fig:ui-overview}.

\subsubsection{Adaptive display of feature-based keyframes (DF5)}
\label{sec:implementation-ui-adaptive-display}

To facilitate the comparison of the feature-based keyframes extracted from each trajectory,
we render keyframe buttons dynamically based on information sent alongside the keyframe to highlight information (upon hover), including the keyframe's start/stop time and an annotated thumbnail.
For features wholly unique to each trajectory, such as ``Collisions'', the buttons are labeled as per the collision number (\ie ``Collision 1'', ``Collision 2'', ``Collision 3'', \etc), and placed underneath (\Cref{fig:ui-collision-hover}).
For features shared between both videos, such as the ``Pick Up Time'' or ``Highest Point'' features, we label buttons appropriately, and on hover, show the keyframe information for both videos side-by-side (\Cref{fig:ui-keyframe-hover}).
Upon clicking these buttons, the system loops the video (if it is a unique keyframe) or two videos (if it is a shared keyframe) between the start and stop time-frames to allow users to compare both features directly,
thereby helping labelers to compare noteworthy differences between the two videos.

\subsubsection{Feature distribution visualization (DF5)}
\label{sec:implementation-ui-feature-distribution}

In addition, we visualize the feature distribution via density area charts to provide more context regarding the whole picture of the feature in the dataset.
For each trajectory we render a density chart of the ``outlying'' feature, \ie the feature with the maximum absolute value of its z-score \cite{curtis2016mystery}, \ie $\lvert\frac{x - \mu}{\sigma}\rvert$.
Additionally, the following values were illustrated and color-coded on the graph with easily identifiable colors for the overall statistics or to match the corresponding trajectories:
\begin{itemize}
    \item the mean $\mu$ with the range of plus and minus half std $\sigma$, \ie $\mu \pm 0.5\sigma$, in transparent green area to compare the given feature values to the average across the dataset,
    \item the outlying feature value in red line to highlight, and
    \item the same feature but from the other trajectory of the pair in blue line for comparison.
\end{itemize}
The interface typically displays two density charts, each highlighting an outlying feature from one trajectory in the pair.
If both trajectories in the pair share the same outlying feature, the interface will only display one density chart for that feature, with two red lines indicating the feature values from the trajectory pair.

\subsubsection{Progress stepper and feedback (DF4)}
\label{sec:implementation-ui-feedback}

Finally, we add a progress stepper (\Cref{fig:ui-overview}) at the bottom of the interface, as well as prompting windows (\Cref{fig:ui-encouraging-message,fig:ui-taking-rest-message}) to provide users with feedback regarding progress.
The progress bar does not explicitly state the number of completed trajectory pairs to prevent overwhelming the labelers with a large number of pairs. 
Instead, it updates to the next step before each attention-checking pair and provides feedback messages from the server to positively reinforce the user's progress.
These components fulfill \textit{DR3} of real-time attention monitoring and feedback provision from the user interface.

%% file: sections/floats/fig-ui-overview.tex
\begin{figure*}
    \centering
    \includegraphics[width=\textwidth]{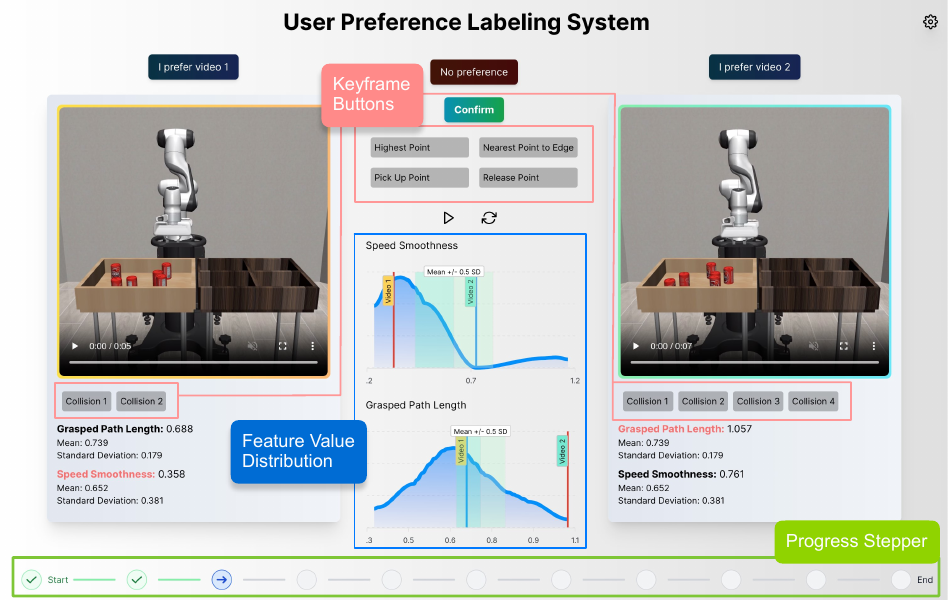}
    \caption{
        The primary component of the user interface is two juxtaposed trajectory videos.
        Labelers can play the videos simultaneously using the play button in the middle or individually using separate controls.
        The labeler can click one button on the top to indicate a preference.
        Auxiliary features include
        looping videos with \textit{keyframe buttons}
        (\Cref{sec:implementation-ui-adaptive-display}, \Cref{fig:ui-collision-hover,fig:ui-keyframe-hover}),
        outlying \textit{feature value distribution}
        (\Cref{sec:implementation-ui-feature-distribution}),
        and a \textit{progress stepper} with messages prompting at each step
        (\Cref{sec:implementation-ui-feedback}, \Cref{fig:ui-encouraging-message,fig:ui-taking-rest-message}).
    }
    \Description{User interface overview. ()}
    \label{fig:ui-overview}
\end{figure*}

%% file: sections/evaluation.tex
\section{Evaluation}
\label{sec:evaluation}

\input{sections/floats/tab-evaluation-metric.tex}

We conducted a between-subjects study to evaluate \tool{}'s effectiveness in improving the quality of preference labels with the Institutional Review Board (IRB) approval from the {University} Research Ethics Committee.

\subsection{Participants}
\label{sec:evaluation-participants}

Forty-two participants
($21$ identified themselves as female, $18$ identified themselves as male, and $2$ identified themselves as non-binary,
in the age range of $19$ to $40$ with a mean of $23.4$ and a standard deviation of $3.2$),
who did not participate in the formative study,
were recruited in the user study.
Their familiarity with the data labeling system ($1$ for Not familiar at all --- $5$ for Extremely familiar) ranges from $1$ to $4$ with a mean of $2.2$ and a standard deviation of $0.9$. Additionally, their familiarity with robotics ($1$ for Not familiar at all --- $5$ for Extremely familiar) ranges from $1$ to $4$ with a mean of $2.0$ and a standard deviation of $0.8$.

\subsection{Task and Condition}
\label{sec:evaluation-task-condition}

\begin{figure*}
    \centering
    \begin{subfigure}{0.49\textwidth}
        \centering
        \includegraphics[width=\textwidth]{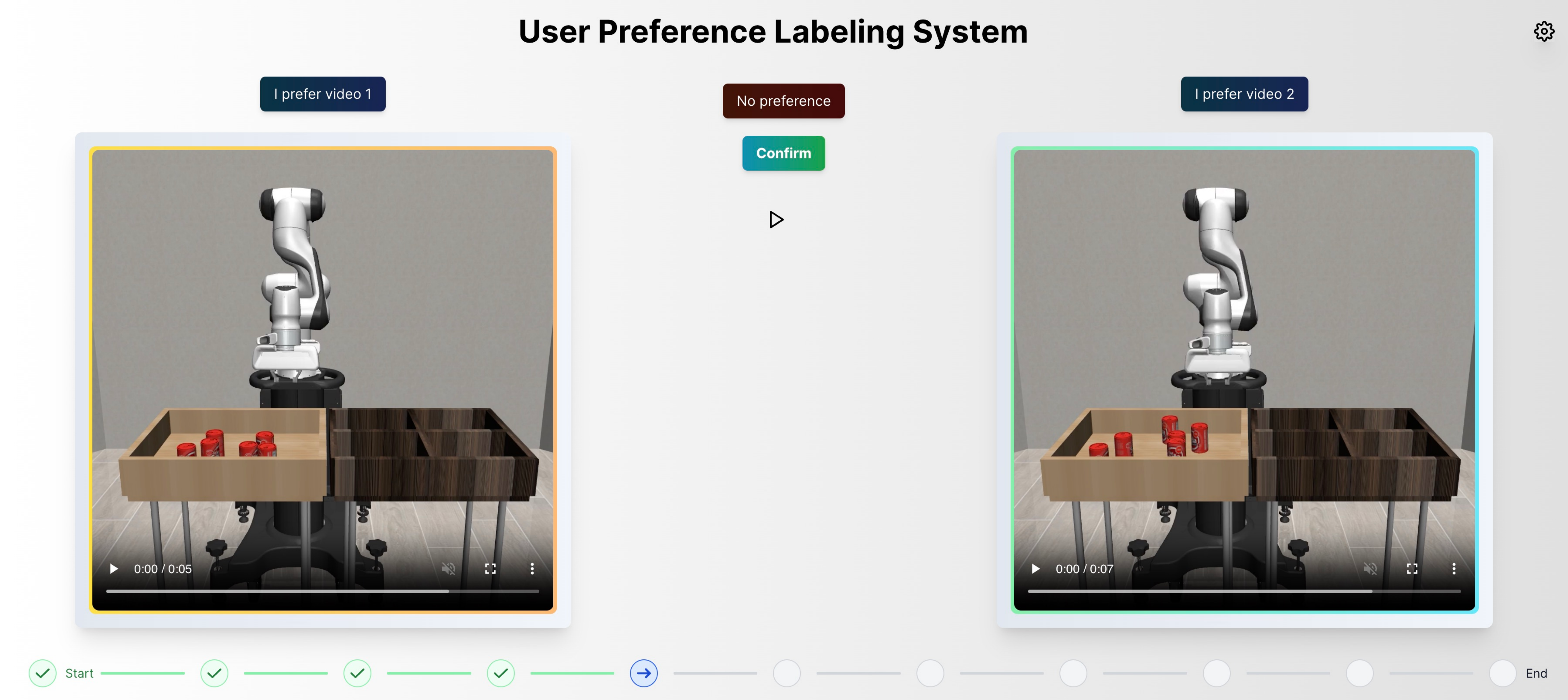}

        \caption{Baseline System}
        \label{fig:ui-baseline}
    \end{subfigure}
    \hfill
    \begin{subfigure}{0.49\textwidth}
        \centering
        \includegraphics[width=\textwidth]{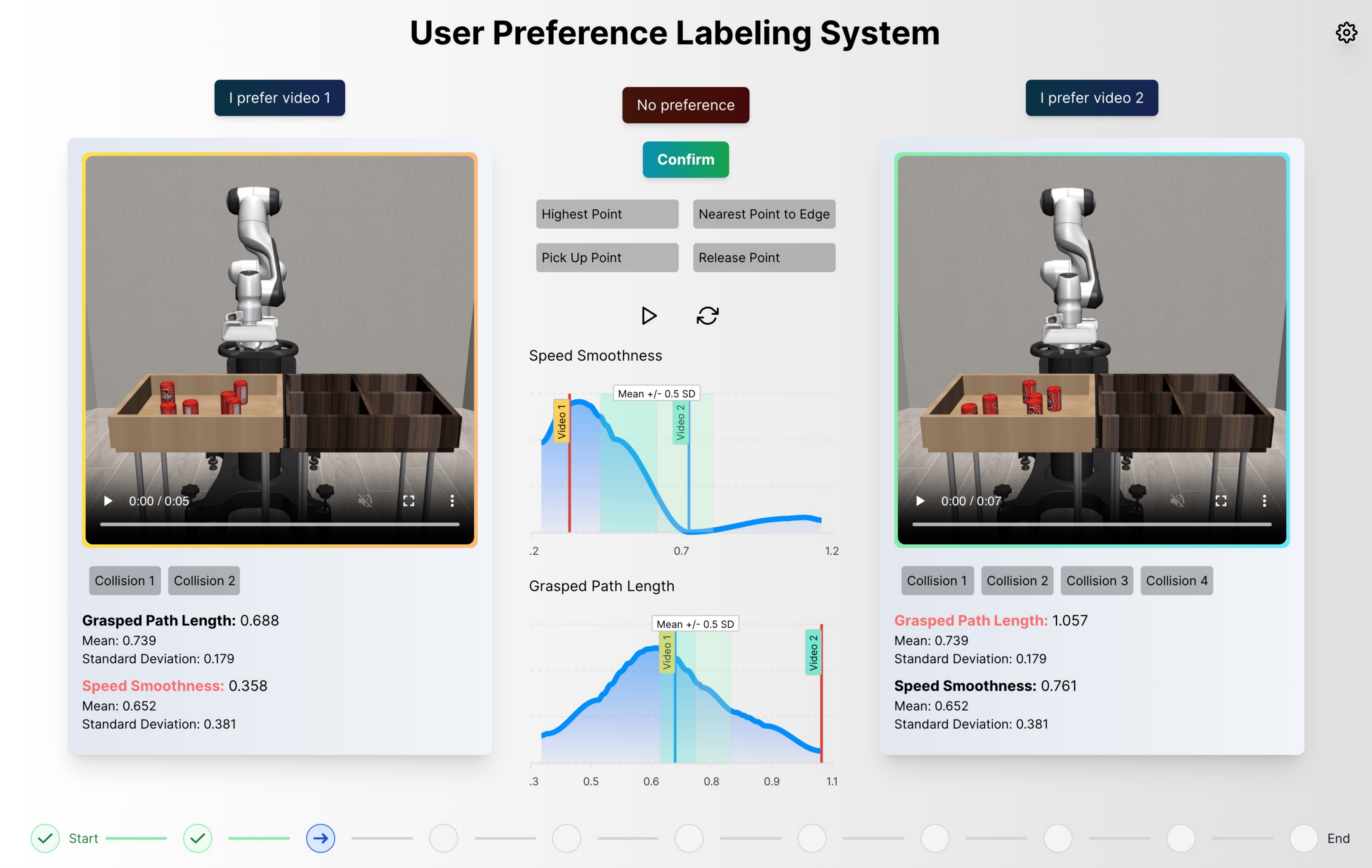}
        \caption{\tool{}}
        \label{fig:ui-farpls}
    \end{subfigure}
    
    \caption{
        System interfaces of two evaluation conditions: Baseline System and \tool{}.
        The Baseline System features two side-by-side videos and a stepped progress bar,
        while \tool{} incorporates additional information, including keyframes, feature distributions, and feedback messages.}
    \Description{The interface of the baseline system and \tool{}.}
    \label{fig:ui-evaluation}
\end{figure*}

We deployed two online trajectory preference labeling systems to a web server for the user study:
a) the baseline system with a conventional interface with two videos side-by-side and the stepper only (\Cref{fig:ui-baseline}), and
b) our proposed system \tool{} with features and keyframes shown as auxiliary information additional to the two videos and the stepper (\Cref{fig:ui-farpls}).
The method of between-subjects design was used,
and we randomly split the participants into two groups, with baseline group (N = 21) and \tool{} group (N = 21).

To prepare data with similar feature distributions with \dataset, we calculated sample weights based on the feature values extracted from each trajectory and stratified sampled $30$ representative samples (\ie ${30 \times 29}/{2} = 435$ unique trajectory pairs) from \dataset{} with the sample weights for the user study.
Our comparative study utilizes these $30$ sampled trajectories to compare the performance of the two systems.
We invited each participant to label $115$ pairs (with $105$ unique pairs from the sampled trajectories and $10$ pairs for consistency checking) of robot pick-and-place trajectories through the system.
In the baseline condition, the system randomly assigns $105$ unique pairs to each participant, at the same time, making sure that at least $5$ participants label each pair.
In the \tool{} condition, \tool{} first clusters the $30$ trajectories using the criterion vector series
and dynamically prompts the trajectory pairs according to the strategy mentioned in \Cref{sec:implementation-server-prompting}, which balances different ranking metrics when guaranteeing label skewness.

Labeling $105$ unique pairs consists of $10$ steps, each step containing $10$ (the number is $15$ for the first step) unique pairs and $1$ consistency checking pair.
At the end of each step, both systems randomly select a consistency-checking pair from the labeled pairs and prompt it to the user.
We did not tell the participants the numbers of unique pairs and consistency-checking pairs.
Still, both systems can show the status (incomplete/active/completed) of $10$ steps in the stepper to the participants and prompt a stop message when all pairs are labeled.

\subsection{Procedure}
\label{sec:evaluation-procedure}

First, the host of the study session provided instructions on the robot task, the labeling system for the corresponding group, and the labeling task and ensured that participants were clear about everything.
Then, the participants were asked to label $115$ pairs of robot pick-and-place trajectories through the system.
Finally, we conducted a post-study survey with the $7$-point Likert scale and open-ended interview questions.
The FARPLS group was administered additional $7$-point Likert scale questions regarding the auxiliary features in FARPLS,
while distinct open-ended interview questions were posed to the two groups.
The whole study session took about $90$ to $120$ minutes, varying depending on the learning, loading, and labeling speed.

\subsection{Evaluation Measurement}
\label{sec:evaluation-measurement}

We collected quantitative measures by the collected preference data, logging participant interactions, and Likert scale questions. We also conducted a semi-structured interview to collect participants' other feedback.
\Cref{tab:evaluation-metric} lists the evaluation metrics used in our study.
The objective metrics include consistency and labeling time.
The subjective metrics include cognitive load, confidence and challenges in \Cref{sec:formative-study-challenges}.
For the \tool{} group, we further explored their perspectives on the auxiliary information provided by the system and their perceptions of how well the design goals were fulfilled (\Cref{tab:metrics-farpls-only}).

Additionally, we conducted a semi-structured interview to collect participants' other feedback.
For both groups, we asked about participants' criteria with features and priorities and their comments and suggestions on the system.
For the baseline group, we also discussed the features and keyframes in \Cref{tab:criteria-features} at the end.
For the \tool{} group, we asked more about their opinions on each design requirement (\Cref{sec:formative-study-requirements}).

%% file: sections/floats/tab-evaluation-metric.tex
\setlength{\colonewidth}{\dimexpr .25\textwidth - 2\tabcolsep}
\setlength{\tabletextwidth}{\dimexpr \textwidth - 2\tabcolsep}
\setlength{\coltwowidth}{\dimexpr \tabletextwidth - \colonewidth - 2\tabcolsep}

\begin{table*}
    \centering
    \caption{Evaluation Metrics for the User Study, including Objective and Subjective Metrics.}
        \begin{tabular}{m{\colonewidth}m{\coltwowidth}}
            \toprule
            \textbf{Objective Metrics}
             &
            \textbf{Definitions}
            \\
            \midrule
            Consistency
             &
            The percentage of consistent labels in the $10$ consistency checking pairs, taking interval values in $\{0, 0.1, \ldots, 0.9, 1\}$
            \\
            \cmidrule{1-2}
            Labeling   Time
             &
            The total labeling time (loading time eliminated), which takes continuous values in $[0, \infty)$
            \\
            \bottomrule
            \toprule
            \textbf{Subjective Metrics}
             &
            \textbf{$7$-Likert Scale Questions}
            \\
            \midrule
            \multicolumn{2}{c}{General Questions}
            \\
            \cmidrule{1-2}
            Cognitive Load
             &
            How mentally challenging was it for you to compare and specify your preference over the video pairs in general? ($1$ - ``Not mentally challenging at all'' to $7$ - ``Extremely mental challenging'')
            \\
            \cmidrule{1-2}
            Confidence
             &
            How confident were you about your labels in general? ($1$ - ``Not confident at all'' to $7$ - ``Extremely confident'')
            \\
            \midrule
            \multicolumn{2}{c}{Questions About Three Challenges}
            \\
            \midrule
            \multicolumn{2}{m{\tabletextwidth}}{Please rate how much you agree with the following statements on a scale of $1$ to $7$, where $1$ is ``strongly disagree'' and $7$ is ``strongly agree'':}
            \\
            \cmidrule{1-2}
            \multirow{6}{\colonewidth}{\textbf{C1} Comparison Criteria}
             &
            \textbf{C1 [criteria establishment]} I can easily establish comparison criteria in general.
            \\
            \cmidrule{2-2}
             &
            \textbf{C1-1 [criteria coverage]} My comparison criteria can cover all the new situations in the later videos.
            \\
            \cmidrule{2-2}
             &
            \textbf{C1-2 [feature coverage]} I am clear about the set of features I rely on to decide the priority of each criterion.
            \\
            \cmidrule{2-2}
             &
            \textbf{C1-3 [feature distribution]} I am clear about the scope of each feature required to determine the priority of each criterion.
            \\
            \cmidrule{1-2}
            \multirow{8}{\colonewidth}{\textbf{C2} Trajectory Details}
             &
            \textbf{C2 [detail overlooking]} I feel that I may overlook some important details when viewing and comparing two trajectories.
            \\
            \cmidrule{2-2}
             &
            \textbf{C2-1 [robotic knowledge]} My knowledge of this robot arm task affects the kind of details I pay attention to.
            \\
            \cmidrule{2-2}
             &
            \textbf{C2-2 [feature support]} The system provides enough support for me to identify features that may be important to this robot arm task in practice.
            \\
            \cmidrule{2-2}
             &
            \textbf{C2-3   [comparison support]} The system provides enough support for me to compare the differences between the two videos.
            \\
            \cmidrule{1-2}
            \multirow{6}{\colonewidth}{\textbf{C3} User Experience}
             &
            \textbf{C3-1} The preference labeling process is \textbf{[easy]}.
            \\
            \cmidrule{2-2}
             &
            \textbf{C3-2} The preference labeling process is \textbf{[boring]}.
            \\
            \cmidrule{2-2}
             &
            \textbf{C3-2} I receive \textbf{[encouragement]} in the preference labeling process.
            \\
            \cmidrule{2-2}
             &
            \textbf{C3-4} I receive \textbf{[feedback]} on my performance in the preference labeling process.
            \\
            \cmidrule{2-2}
             &
            \textbf{C3-5} I find the preference labeling process \textbf{[rewarding]}.
            \\
            \bottomrule
        \end{tabular}
        \label{tab:evaluation-metric}
\end{table*}

\begin{table*}
    \centering
    \caption{Subjective Questions for \tool{} Group Only}
    \begin{tabular}{m{\colonewidth}m{\coltwowidth}}
        \toprule
        \textbf{Metrics}
         &
        \textbf{$7$-Likert Scale Questions}
        \\
        \midrule
        \multicolumn{2}{m{\tabletextwidth}}{Please rate how much you agree with the following statements about the auxiliary information provided by the system on a scale of $1$ to $7$, where $1$ is ``strongly disagree'' and $7$ is ``strongly agree'':}
        \\
        \cmidrule{1-2}
        \multirow{9}{\colonewidth}{Auxiliary Information}
         &
        \textbf{AX1} The auxiliary information provided by the system is \textbf{[informative]}.
        \\
        \cmidrule{2-2}
         &
        \textbf{AX2} The auxiliary information provided by the system is \textbf{[relevant]}.
        \\
        \cmidrule{2-2}
         &
        \textbf{AX3} The auxiliary information provided by the system is helpful for establishing preference \textbf{[criteria]}.
        \\
        \cmidrule{2-2}
         &
        \textbf{AX4} The auxiliary information provided by the system prevents me from observing more \textbf{[details]} in the videos.
        \\
        \cmidrule{2-2}
         &
        \textbf{AX5} The auxiliary information provided by the system is helpful for \textbf{[comparing]} the differences between the two videos.
        \\
        \cmidrule{2-2}
         &
        \textbf{AX6} The auxiliary information provided by the system is \textbf{[overwhelming]}.
        \\
        \cmidrule{2-2}
         &
        \textbf{AX7} The auxiliary information provided by the system is \textbf{[distracting]}.
        \\
        \cmidrule{1-2}
        \multicolumn{2}{m{\tabletextwidth}}{We will present you with three designs of this system. Please rate how helpful each design is in a particular aspect on a scale of $1$ to $7$, where $1$ is ``not at all helpful'' and $7$ is ``very helpful'':}
        \\
        \cmidrule{1-2}
        \multicolumn{2}{m{\tabletextwidth}}{1. We group the trajectories according to a set of features illustrated in \Cref{tab:criteria-features}. Subsequently, we present the trajectories with greater variations in these features to you initially.}
        \\
        \cmidrule{1-2}
        \multirow{4}{\colonewidth}{\textbf{DR1} Prompting Strategy}
         &
        \textbf{DR1C1-1 [initial familiarity]} How helpful is this design in facilitating familiarity with diverse situations and establishing criteria for comparison?
        \\
        \cmidrule{2-2}
         &
        \textbf{DR1C2-2 [detail aware]} How helpful is this design in improving your perception of the details of the trajectory, regardless of your level of robotics expertise?
        \\
        \cmidrule{1-2}
        \multicolumn{2}{m{\tabletextwidth}}{2. This system presents the distributions of representative trajectory features and keyframes from a particular view.}
        \\
        \cmidrule{1-2}
        \multirow{4}{\colonewidth}{\textbf{DR2} Feature and Keyframe}
         &
        \textbf{DR2C1-2 [criteria priority]} How helpful is this design in assisting your comprehension of each feature's range to determine the priority of your criteria?
        \\
        \cmidrule{2-2}
         &
        \textbf{DR2C2 [sense making]} How helpful is this design in enhancing your understanding of this robot task and gaining access to more information?
        \\
        \cmidrule{1-2}
        \multicolumn{2}{m{\tabletextwidth}}{3. This system offers a real-time attention-monitoring feature. If you label inconsistently, the system will prompt you to take a break. On the other hand, if you label consistently, it will motivate you to continue.}
        \\
        \cmidrule{1-2}
        \multirow{2}{\colonewidth}{\textbf{DR3} Attention Monitoring and Feedback}
         &
        \textbf{DR3C3-2} How helpful is this design in \textbf{[decreasing the boredom]} of labeling preferences?
        \\
        \cmidrule{2-2}
         &
        \textbf{DR3C3-5} How helpful is this design in \textbf{[increasing the reward]} of the preference labeling process?
        \\
        \bottomrule
    \end{tabular}
    \label{tab:metrics-farpls-only}
\end{table*}

%% file: sections/result.tex
\section{Results}
\label{sec:results}

To answer RQ3, we comprehensively analyze all the metrics for each group.
We first conduct significant tests for each metric in two-sided, and then for significant metrics, we conduct additional one-sided tests.
We also answer how well \tool{} solves the challenges by each design requirement and the participants' feedback on the design requirements.
We further highlight the participants' most insightful comments and suggestions.

\begin{figure*}
    \centering
    \begin{subfigure}{.49\textwidth}
        \centering
        \includegraphics[height=.9\textwidth]{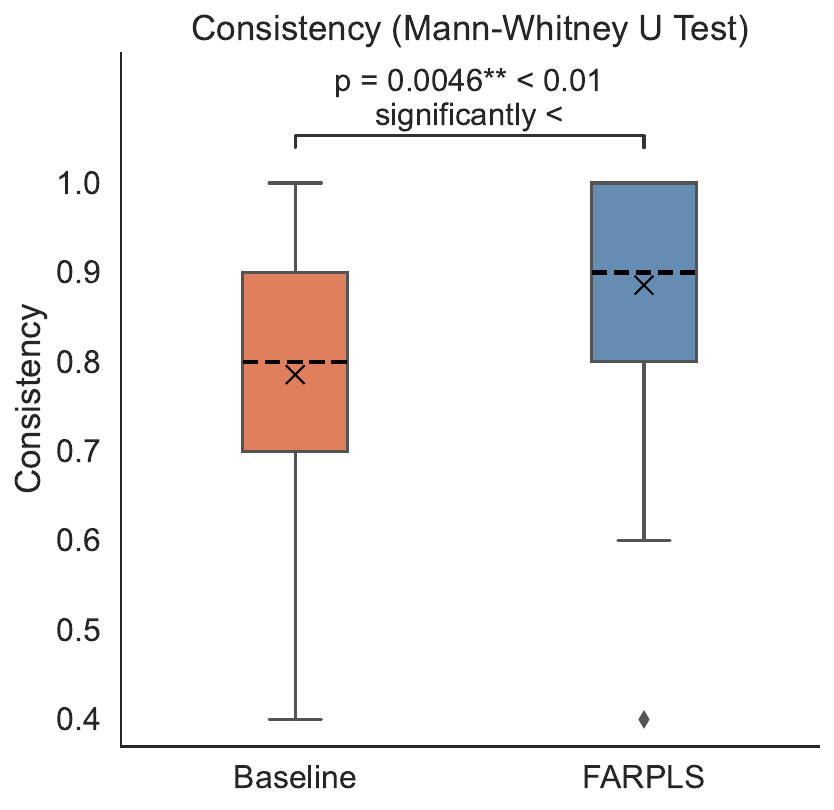}
        \subcaption{The boxplot of \textbf{Consistency} of two groups}
        \label{fig:results-consistency}
    \end{subfigure}
    \hfil
    \begin{subfigure}{.49\textwidth}
        \centering
        \includegraphics[height=.9\textwidth]{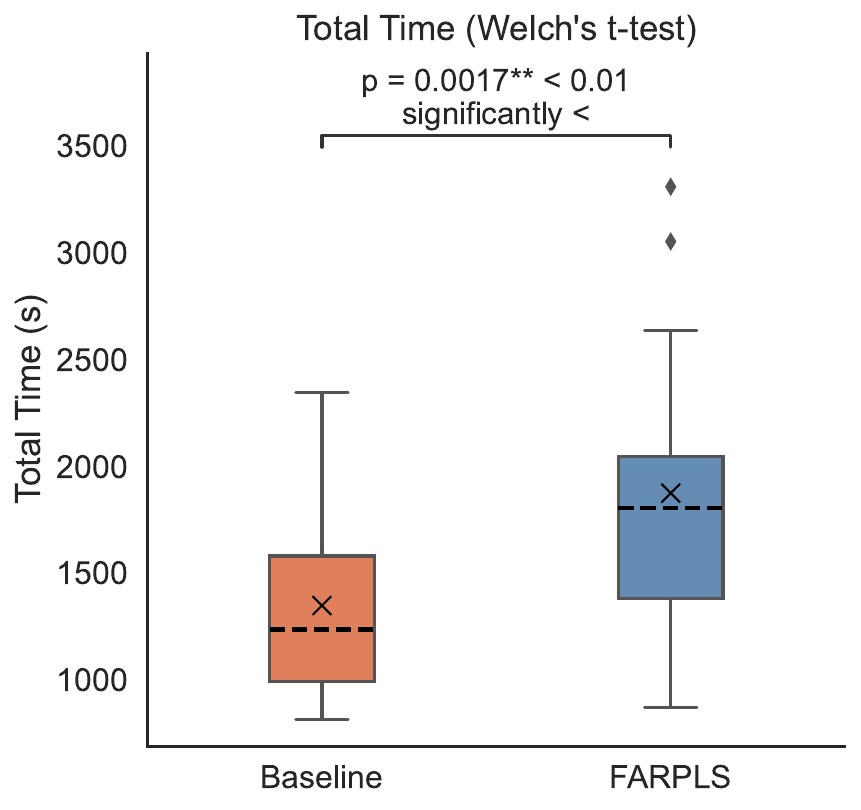}
        \subcaption{The boxplot for \textbf{Total Time} of two groups.}
        \label{fig:results-time}
    \end{subfigure}
    
    \caption{The distributions of two subjective metrics \textbf{Consistency} and \textbf{Total Time} in boxplots comparing two system conditions.}
    \Description{Results of the participants.}
    \label{fig:results-objective}
\end{figure*}

\subsection{Consistency}
\label{sec:results-consistency}

The \tool{} group has a significantly higher consistency score than the baseline group according to the Mann-Whitney U test \cite{mann1947test}, $U = 321.5$, $p = 0.0046\text{**} < 0.01$.
The participants in the baseline group have an average consistency score of $0.79$ ($SD = 0.15$), and the participants in the \tool{} group have an average consistency score of $0.89$ ($SD = 0.16$).
\Cref{fig:results-consistency} shows the distribution of consistency scores in boxplots for both groups.
The result indicates that participants are more consistent with the help of \tool{}.

\subsection{Labeling Time}
\label{sec:results-time}
\subsubsection{Total Labeling Time}
\label{sec:results-total-time}

Participants spend significantly more time labeling with \tool{} than with the baseline system according to Welch's t-test, $t(36.18) = 3.127$, $p = 0.0017\text{**} < 0.01$, $95\% ~CI~ [242.36, \infty]$.
On average, the participant in the baseline group spent $1350$s ($SD = 448$) labeling all the trajectories, while the participant in the \tool{} group spent $1877$s ($SD = 628$).
\Cref{fig:results-time} shows the distribution of total labeling time spent by two groups of participants in boxplots.
The result is within expectations, since participants in the \tool{} group need to spend more time observing the auxiliary information in \tool.

\subsubsection{Learning Curve}
\label{sec:results-learning-curve}

\begin{figure*}
    \centering
    \includegraphics[width=\textwidth]{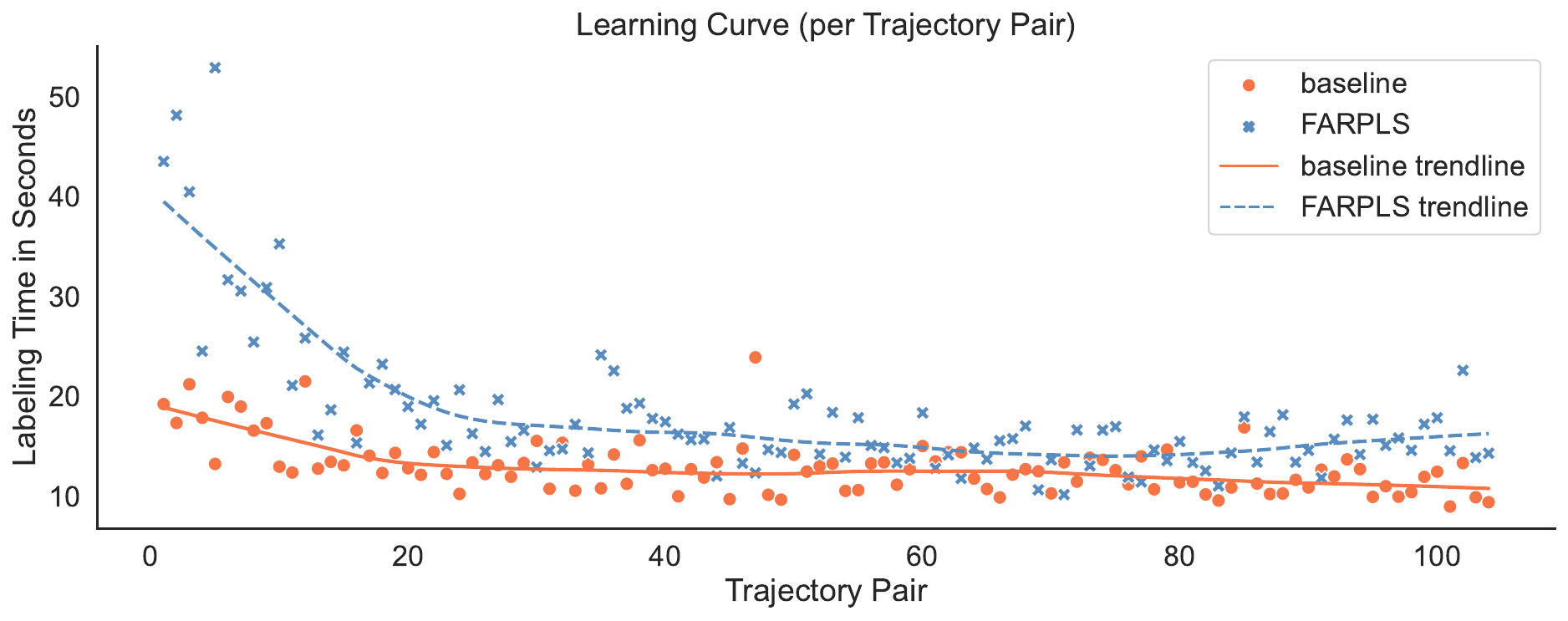}
    
    \caption{{Learning curve showing the participants' average labeling time per pair.}}
    \Description{Learning curve of the participants.}
    \label{fig:learning-curve}
\end{figure*}

We plot a learning curve to analyze how the participants' labeling time for each trajectory pair changes during the experiment in~\Cref{fig:learning-curve}.
The fitted smoothing lines (via Locally Weighted Scatterplot Smoothing (LOESS) method \cite{cleveland1979robust}) show the trend of the average labeling time per pair of participants.
The average labeling time per pair of participants decreases as the experiment progresses,
which indicates that the participants are getting more familiar with the labeling task and their comparison criteria.
A sharper decreasing trend in the \tool{} group than in the baseline group also indicates that the participants in the \tool{} group are learning faster than the participants in the baseline group.
There is also a slight increase at the end of the \tool{} group's learning curve, possibly due to the prompted pairs' increased difficulty.

\subsection{Cognitive Load and Confidence}
\label{sec:results-cognitive-load-confidence}

\begin{figure*}
    \centering
    \includegraphics[width=0.52\textwidth]{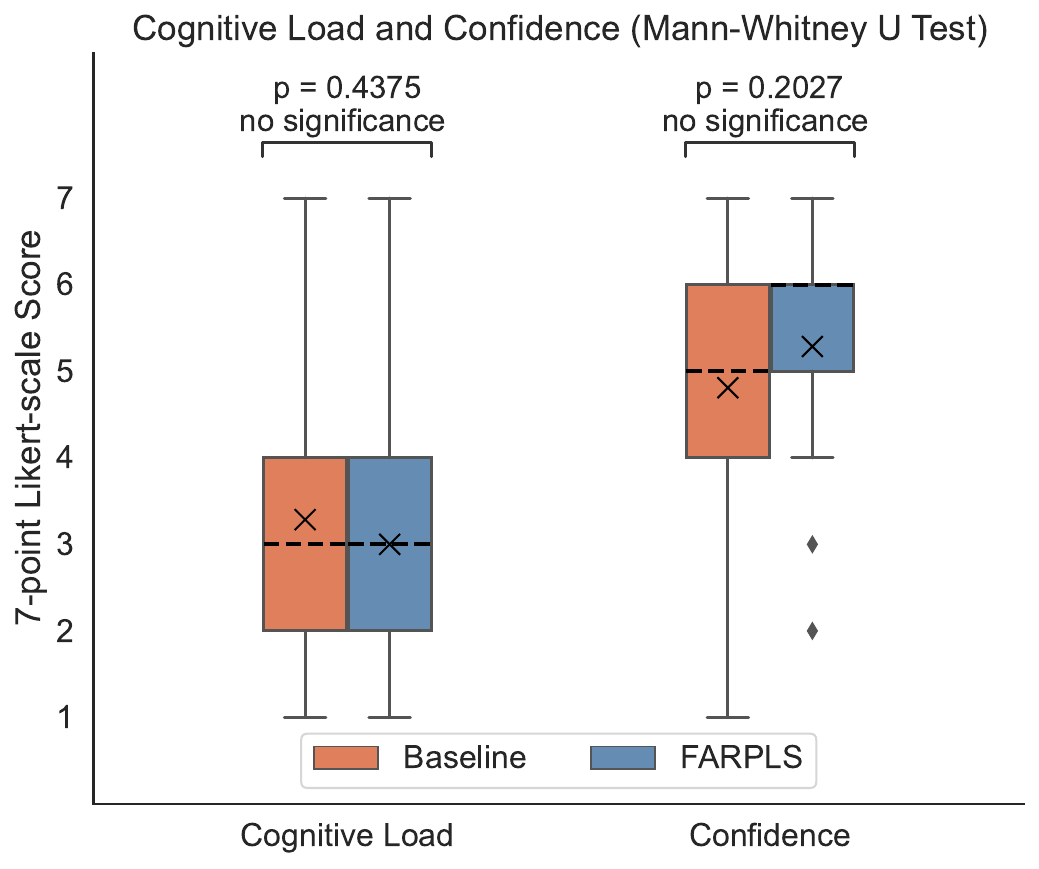}
    
    \caption{The distributions of \textbf{Cognitive Load} and \textbf{Confidence} in boxplots comparing two system conditions.}
    \Description{Cognitive Load and Confidence}
    \label{fig:results-cognitive-load-confidence}
\end{figure*}

\Cref{fig:results-cognitive-load-confidence} shows the distributions of the ratings of two subjective metrics, {Cognitive Load} and {Confidence}.

\subsubsection{Cognitive Load}
\label{sec:results-cognitive-load}
There is no significant difference in the cognitive load scores between the two groups according to the Mann-Whitney U test, $U = 190.0$, $p = 0.4375 > 0.05$.
The participants in the baseline group have an average cognitive load score of $3.28$ ($SD = 1.42$), and the participants in the \tool{} group have an average cognitive load score of $3.00$ ($SD = 1.64$).
This result indicates that the additional design features in \tool{} do not affect the participants' cognitive load.

\subsubsection{Confidence}
\label{sec:results-confidence}
There is no significant difference in the confidence scores between the two groups according to the Mann-Whitney U test, $U = 166.0$, $p = 0.2027 > 0.05$.
The participants in the baseline group have an average confidence score of $4.81$ ($SD = 1.40$), and the participants in the \tool{} group have an average confidence score of $5.29$ ($SD = 1.16$).
The participants are generally confident in their labeling according to the means ($> 4$) and the plots in \Cref{fig:results-cognitive-load-confidence}.
However, the baseline group's {Consistency} metric is significantly lower than the \tool{} group.
This finding suggests that participants may have overconfidence in their self-assessments of the {Confidence} metric.

\subsection{Challenges and Design Requirements}
\label{sec:results-challenges-design-requirements}

\Cref{tab:results-challenges-statistics} shows the statistics for metrics and \Cref{fig:results-c1-boxplot,fig:results-c2-boxplot,fig:results-c3-boxplot} show the distributions of the metrics.

\begin{table*}
    \centering
    \caption{{Challenge Metrics Statistics and Mann-Whitney U Test Results}}
    
    \begin{tabular}{lllllllll}
        \toprule
                                     &
        \multirow{2}{*}{{Metrics}}   &
        \multicolumn{2}{c}{Baseline} &
        \multicolumn{2}{c}{\tool}    &
        \multicolumn{3}{c}{\begin{tabular}[c]{@{}c@{}}Mann-Whitney U Test \\ (\tool v.s. Baseline)\end{tabular}} \\ \cmidrule(lr){3-4} \cmidrule(lr){5-6} \cmidrule(lr){7-9}
                                     &
                                     &
        {mean}                       &
        {std}                        &
        {mean}                       &
        {std}                        &
        {U-value}                    &
        {alt. hypo.}                 &
        {p-value}
        \\ \midrule
        C1                           &
        [criteria establishment]     &
        {$4.67$}                     &
        {$1.59$}                     &
        {$5.62$}                     &
        {$0.86$}                     &
        {$295.5$}                    &
        {greater}                    &
        {$0.0246$*}
        \\
        C1-1                         &
        [criteria coverage]          &
        {$3.86$}                     &
        {$1.77$}                     &
        {$5.05$}                     &
        {$1.20$}                     &
        {$307.5$}                    &
        {greater}                    &
        {$0.0128$*}
        \\
        C1-2                         &
        [feature coverage]           &
        {$4.81$}                     &
        {$1.54$}                     &
        {$5.57$}                     &
        {$1.03$}                     &
        {$281.0$}                    &
        {two-sided}                  &
        {$0.1173$}
        \\
        C1-3                         &
        [feature distribution]       &
        {$4.67$}                     &
        {$1.46$}                     &
        {$5.43$}                     &
        {$0.98$}                     &
        {$275.0$}                    &
        {two-sided}                  &
        {$0.1533$}
        \\ \midrule
        C2                           &
        [detail overlooking]         &
        {$4.00$}                     &
        {$1.70$}                     &
        {$3.90$}                     &
        {$1.41$}                     &
        {$214.5$}                    &
        {two-sided}                  &
        {$0.8870$}
        \\
        C2-1                         &
        [robotic knowledge]          &
        {$3.81$}                     &
        {$1.91$}                     &
        {$3.43$}                     &
        {$1.80$}                     &
        {$193.0$}                    &
        {two-sided}                  &
        {$0.4901$}
        \\
        C2-2                         &
        [feature support]            &
        {$4.10$}                     &
        {$1.73$}                     &
        {$5.76$}                     &
        {$0.94$}                     &
        {$349.0$}                    &
        {greater}                    &
        {$0.0005$***}
        \\
        C2-3                         &
        [comparison support]         &
        {$4.76$}                     &
        {$1.76$}                     &
        {$5.86$}                     &
        {$1.20$}                     &
        {$306.5$}                    &
        {greater}                    &
        {$0.0125$*}
        \\ \midrule
        C3-1                         &
        [easy]                       &
        {$4.71$}                     &
        {$1.95$}                     &
        {$5.33$}                     &
        {$1.43$}                     &
        {$256.0$}                    &
        {two-sided}                  &
        {$0.3643$}
        \\
        C3-2                         &
        [boring]                     &
        {$4.95$}                     &
        {$1.63$}                     &
        {$4.19$}                     &
        {$1.36$}                     &
        {$152.5$}                    &
        {less}                       &
        {$0.0419$*}
        \\
        C3-3                         &
        [encouragement]              &
        {$3.62$}                     &
        {$1.60$}                     &
        {$4.57$}                     &
        {$1.54$}                     &
        {$288.0$}                    &
        {greater}                    &
        {$0.0424$*}
        \\
        C3-4                         &
        [feedback]                   &
        {$2.95$}                     &
        {$1.56$}                     &
        {$4.81$}                     &
        {$2.02$}                     &
        {$334.0$}                    &
        {greater}                    &
        {$0.0019$**}
        \\
        C3-5                         &
        [rewarding]                  &
        {$3.48$}                     &
        {$1.83$}                     &
        {$4.71$}                     &
        {$1.23$}                     &
        {$312.0$}                    &
        {greater}                    &
        {$0.0101$*}
        \\ \bottomrule
    \end{tabular}
    \label{tab:results-challenges-statistics}
\end{table*}

\subsubsection{C1 - Difficulty in forming criteria}
\label{sec:results-c1}

\begin{figure*}
    \centering
    \includegraphics[width=0.75\textwidth]{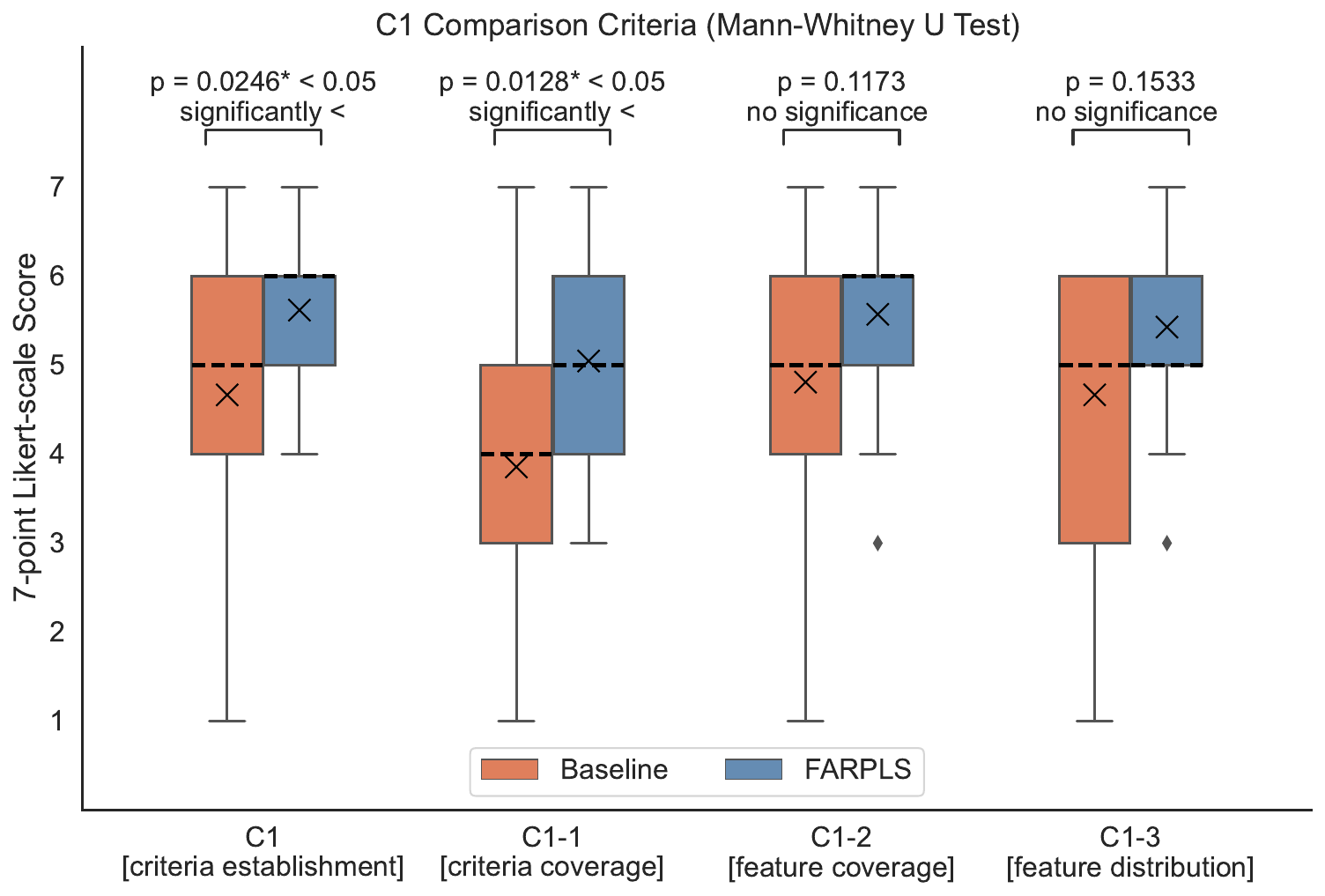}
    
    \caption{The distributions of metrics about \textbf{C1} (Comparison Criteria) in boxplots comparing two system conditions.}
    \Description{The distributions of metrics about \textbf{C1} (Comparison Criteria) in boxplots comparing two system conditions.}
    \label{fig:results-c1-boxplot}
\end{figure*}

\Cref{fig:results-c1-boxplot} shows the distributions of metrics about \textit{C1 - Difficulty in forming criteria} in boxplots comparing two system conditions. According to the statistics in \Cref{tab:results-challenges-statistics}, we have the following results:

\paragraph{C1 [criteria establishment]}
This metric in \tool{} group is significantly higher than that in the baseline group according to the Mann-Whitney U test.
Participants in the \tool{} group find it significantly easier to establish criteria in general compared to the baseline group.

\paragraph{C1-1 [criteria coverage]}
This metric in \tool{} group is significantly higher than that in the baseline group according to the Mann-Whitney U test.
Participants in the \tool{} group reckon their criteria can cover more new situations in the later videos than those in the baseline group.

\paragraph{C1-2 [feature coverage]}
This metric has no significant difference between the two groups according to the Mann-Whitney U test.
The results show that participants in both groups reckon they are clear about the features to decide the priority of each criterion.

\paragraph{C1-3 [feature distribution]}
This metric has no significant difference between the two groups according to the Mann-Whitney U test.
The results show that participants in both groups reckon they are clear about the scope of features to determine the priority of each criterion.

\paragraph{Summary}
The significant results in C1 and C1-1 (\Cref{tab:results-challenges-statistics}) show that \tool{} can successfully mitigate the challenges in forming criteria.

\subsubsection{C2 - Overlooking trajectory details}
\label{sec:results-c2}

\begin{figure*}
    \centering
    \includegraphics[width=0.73\textwidth]{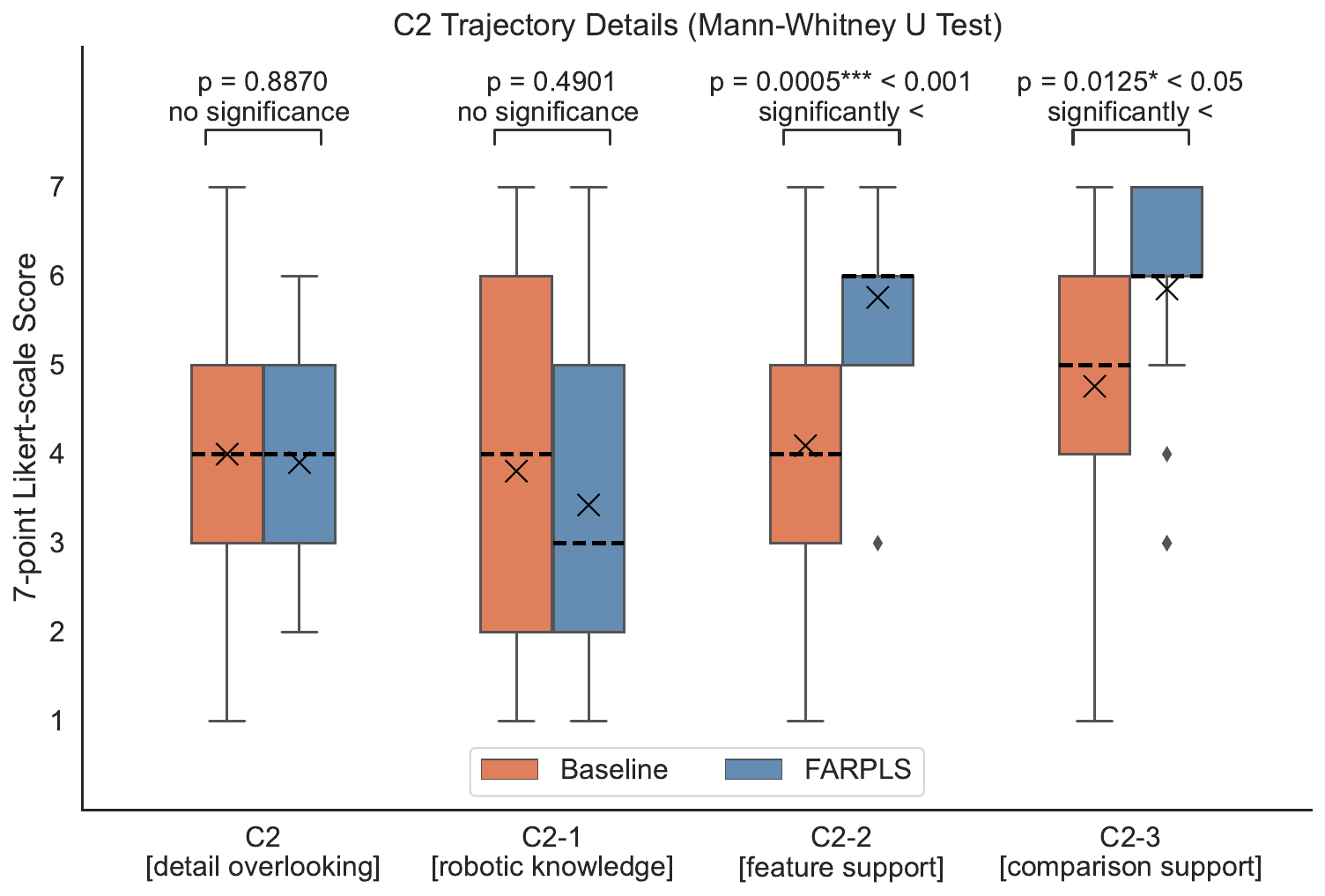}
    
    \caption{The distributions of metrics about \textbf{C2} (Trajectory Details) in boxplots comparing two system conditions.}
    \Description{C2}
    \label{fig:results-c2-boxplot}
\end{figure*}

\Cref{fig:results-c2-boxplot} shows the distributions of metrics about \textit{C2 - Overlooking trajectory details} in boxplots comparing two system conditions. According to the statistics in \Cref{tab:results-challenges-statistics}, we have the following results:

\paragraph{C2 [detail overlooking]}
This metric has no significant difference between the two groups according to the Mann-Whitney U test.
The neutral results of the metric C2 in both groups (both means $\sim 4.00$ and both medians $= 4$) indicate that people do not know whether they overlook trajectory details.

\paragraph{C2-1 [robotic knowledge]}
This metric has no significant difference between the two groups according to the Mann-Whitney U test.
Considering that the two groups' mean metric scores for {C2-1} ($3.81$ and $3.43$) are $< 4.00$,
we can infer that although with large disagreements ($SD = 1.91$ and $=1.80$), many participants in both groups think their knowledge of robotics does not affect their labeling process.

\paragraph{C2-2 [feature support]}
This metric in \tool{} group is significantly higher than that in the baseline group according to the Mann-Whitney U test, $U = 349.0$, $p = 0.0005\text{***} < 0.001$.
Participants in the \tool{} group feel significantly more support than those in the baseline group to identify features that may be important to the robot arm task in practice.

\paragraph{C2-3 [comparison support]}
This metric in \tool{} group is significantly higher than that in the baseline group according to the Mann-Whitney U test.
Participants in the \tool{} group feel significantly more support than those in the baseline group to compare the differences between the two videos.

\paragraph{Summary}
The significant improvements in C2-2 and C2-3 (\Cref{tab:results-challenges-statistics}) show that \tool{} can successfully assist labelers in identifying more features and easily eliciting their preferences.

\subsubsection{C3 - Difficulty in maintaining focus}
\label{sec:results-c3}

\begin{figure*}
    \centering
    \includegraphics[width=0.88\textwidth]{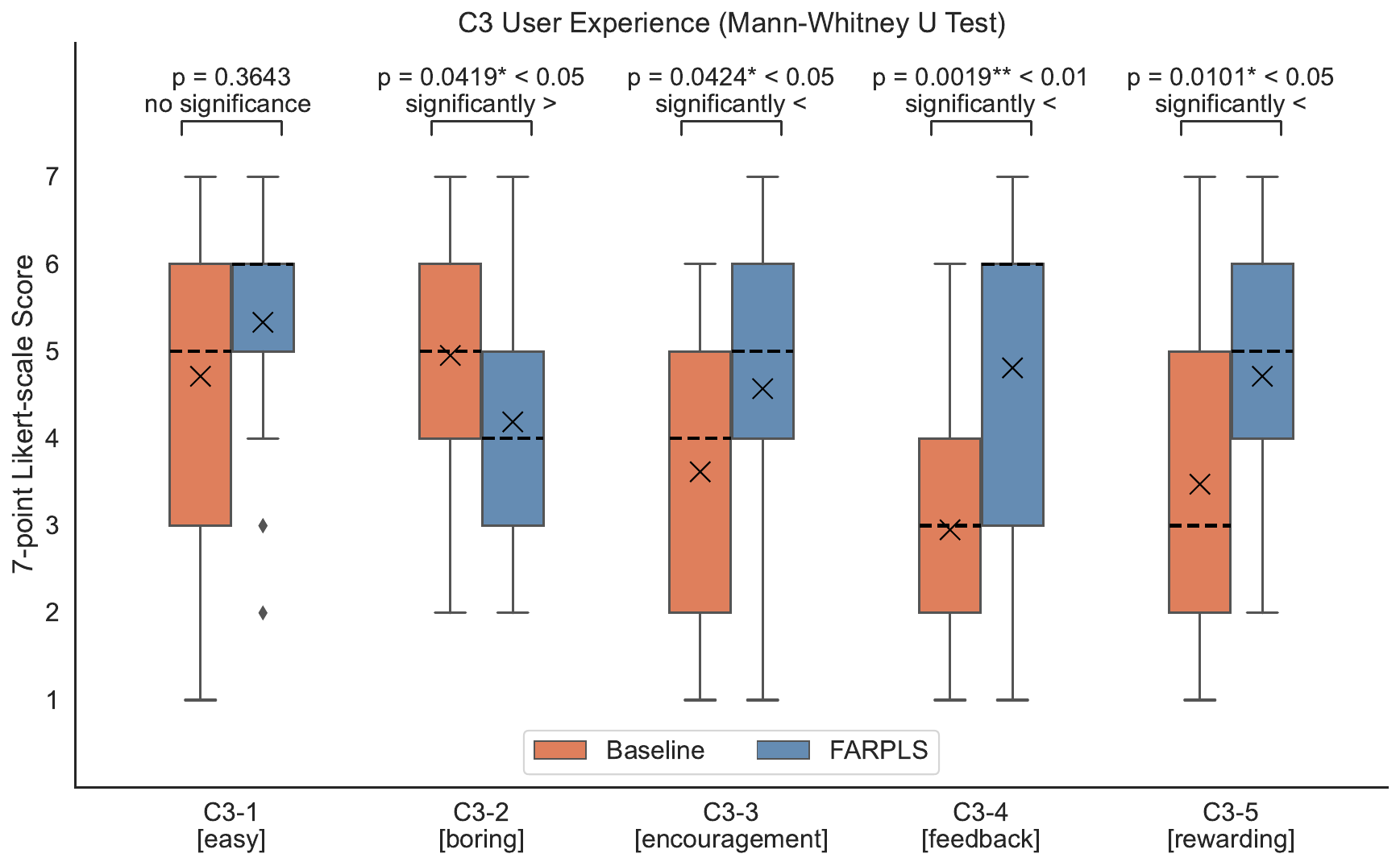}
    
    \caption{The distributions of metrics about \textbf{C3} (User Experience) in boxplots comparing two system conditions.}
    \Description{C3}
    \label{fig:results-c3-boxplot}
\end{figure*}

\Cref{fig:results-c3-boxplot} shows the distributions of metrics about \textit{C3 - Difficulty in maintaining focus} in boxplots comparing two system conditions. According to the statistics in \Cref{tab:results-challenges-statistics}, we have the following results:

\paragraph{C3-1 [easy]}
This metric has no significant difference between the two groups according to the Mann-Whitney U test.
The scores of {C3-1} are high in means ($4.71$ and $5.33 > 4.00$) and medians ($5$ and $6 > 4$) for both groups.
The results show that participants in both groups reckon the labeling task easy.

\paragraph{C3-2 [boring]}
This metric in \tool{} group is significantly lower than that in the baseline group according to the Mann-Whitney U test.
Participants in the baseline group find the labeling process more boring than those in the \tool{} group.

\paragraph{C3-3 [encouragement]}
This metric in \tool{} group is significantly higher than that in the baseline group according to the Mann-Whitney U test.
Participants in the \tool{} group reckon they receive more encouragement in the labeling task than those in the baseline group.

\paragraph{C3-4 [feedback]}
This metric in \tool{} group is significantly higher than that in the baseline group according to the Mann-Whitney U test.
Participants in the \tool{} group think they receive more feedback on their labeling performance than those in the baseline group.

\paragraph{C3-5 [rewarding]}
This metric in \tool{} group is significantly higher than that in the baseline group according to the Mann-Whitney U test.
Participants in the \tool{} group feel the labeling process more rewarding than those in the baseline group.

\paragraph{Summary}
The significant improvements (\Cref{tab:results-challenges-statistics}) in C3-2, C3-3, C3-4 and C3-5 show that the participants in the \tool{} group find the labeling process less tedious and more engaging than those in the baseline group.
The insignificant result of C3-1 [easy] indicates that the participants think the labeling task is easy, resulting in their confidence in their criteria (aligned with the {Confidence} metric)
and their belief in the comprehensiveness of their feature criteria and their familiarity with the distribution of features (aligned with the metrics {C1-2} [feature coverage] and {C1-3} [feature distribution]).
As a result, participants do not think their robotic knowledge affects the details they pay attention to (aligned with the metric {C2-1} [robotic knowledge]).
Therefore, the metrics {C3-1}, {C1-2}, {C1-3}, and {C2-1} are logically aligned with Confidence.
We will discuss the implications of this finding in the discussion section.

\subsubsection{Design requirements}
\label{sec:results-design-requirements}
\begin{figure*}
    \centering
    \includegraphics[width=\textwidth]{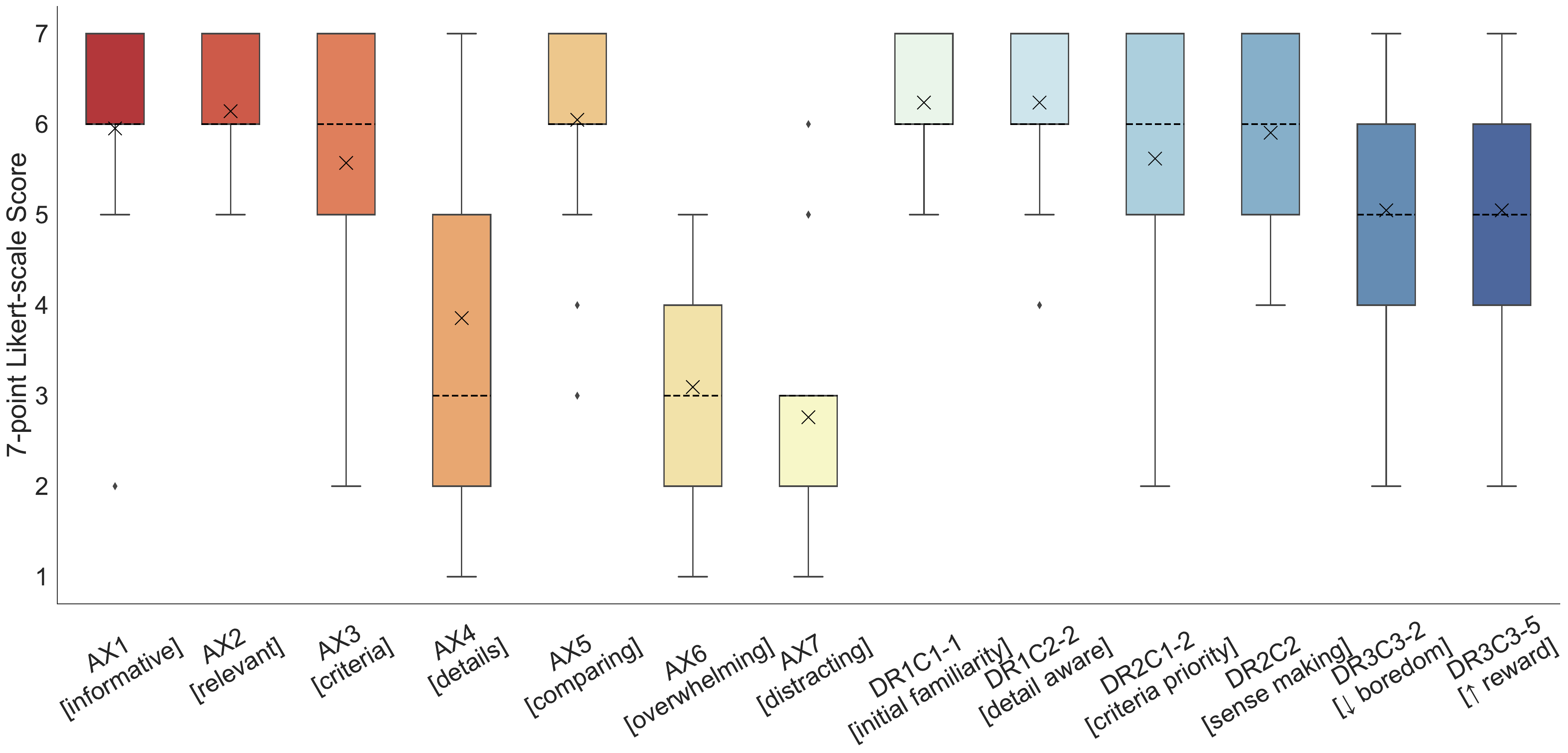}
    
    \caption{{Distributions of additional metrics about auxiliary information and design requirements of the \tool{} group in boxplots.}}
    \Description{Additional questions about design requirements}
    \label{fig:results-additional-questions}
\end{figure*}

The additional metrics for the \tool{} group are for participants to rate the auxiliary information and the helpfulness of design requirements of \tool{},
reported in \Cref{tab:results-additional-questions} and \Cref{fig:results-additional-questions}.
As we can see from the results (relevant means and medians comparing with $4$),
the participants find the auxiliary information informative, relevant, and helpful for establishing comparison criteria.
Moreover, the auxiliary information does not prevent the participants from observing more details in the videos and is helpful for them to compare the trajectories,
neither overwhelming nor distracting.

The rest questions are designed to check whether the design requirements of \tool{} solve the challenges mentioned in \Cref{sec:formative-study}.
We dynamically arranged the order to balance labeling difficulty in the case of {DR1}.
Our prompting strategy during the initial stage significantly facilitated participants in quickly gaining familiarity with the labeling task and establishing their comparison criteria ({DR1C1-1}, $M = 6.24$).
Furthermore, it enabled them to comprehend the intricate details of the trajectories ({DR1C2-2}, $M = 6.24$).
For {DR2},
the participants find the design helpful for them to understand the feature value distributions to determine the priorities ({DR2C1-2}, $M = 5.62$)
and access more information to make sense of the robot task ({DR2C2}, $M = 5.90$).
As shown in the metric results of {DR3},
the participants find the design moderately helpful in decreasing the boredom ({DR3C3-2}, $M = 5.05$)
and increasing the reward ({DR3C3-5}, $M = 5.05$).
Therefore, participants find all the design requirements of \tool{} helpful for the corresponding challenges connected to the DRs in \Cref{fig:C-DR-DF}.
\begin{table*}
    \centering
    \caption{Statistics of additional metrics about auxiliary information and design requirements of the \tool{} group.}
    
    \begin{tabular}{lccccccc}
        \toprule
             & {AX1}         & {AX2}      & {AX3}      & {AX4}     & {AX5}       & {AX6}          & {AX7}
        \\
             & [informative] & [relevant] & [criteria] & [details] & [comparing] & [overwhelming] & [distracting]
        \\ \midrule
        mean & $5.95$        & $6.14$     & $5.57$     & $3.86$    & $6.05$      & $3.10$         & $2.76$
        \\
        std  & $1.12$        & $0.73$     & $1.40$     & $2.13$    & $1.12$      & $1.41$         & $1.30$
        \\ \bottomrule
    \end{tabular}
    \medbreak
    \begin{tabular}{ccccccc}
        \toprule
             & \multicolumn{2}{c}{{DR1} Prompting Strategy}
             & \multicolumn{2}{c}{{DR2} Feature and Keyframe}
             & \multicolumn{2}{c}{{DR3} Attention Monitoring and Feedback}
        \\
        \cmidrule(lr){2-3} \cmidrule(lr){4-5} \cmidrule(lr){6-7}
             & {DR1C1-1}                                                          & {DR1C2-2} & {DR2C1-2} & {DR2C2} & {DR3C3-2} & {DR3C3-5}
        \\ \midrule
        mean & 6.24                                                               & 6.24      & 5.62      & 5.90    & 5.05      & 5.05
        \\
        std  & 0.70                                                               & 0.77      & 1.50      & 0.94    & 1.47      & 1.53
        \\ \bottomrule
    \end{tabular}
    \label{tab:results-additional-questions}
\end{table*}

\subsection{Semi-structured Interview}
\label{sec:results-interview}

\subsubsection{Features in participants' comparison criteria}
\label{sec:results-interview-features-comparison-criteria}

We compared the number of features noticed by participants from two groups. We clarified subtle distinctions in our feature categorization to the participants and tallied their explicit indications of whether they noticed each of them.
``Power Usage'', ``Contact Force'', and ``Orientation'' are the three least noticed features with $18$, $14$, and $8$ participants out of $21$ neglecting them, respectively, as is shown in \Cref{tab:baseline-feature-detect}.
\begin{table}
    \centering
    \caption{Number of baseline group participants \textbf{overlooking} each feature.}
    
    \begin{tabular}{lc}
        \toprule
        Feature               & Number of Participants \\ \midrule
        Collision             & 0                      \\
        Distance              & 2                      \\
        Contact Force         & 14                     \\ \cmidrule(lr){1-2}
        Speed                 & 7                      \\
        Path Length           & 6                      \\
        Time                  & 6                      \\
        Power Usage           & 18                     \\ \cmidrule(lr){1-2}
        Speed Smoothness      & 7                      \\
        Trajectory Smoothness & 2                      \\
        Orientation           & 8                      \\
        Grasp Position        & 7                      \\ \bottomrule
    \end{tabular}
    \label{tab:baseline-feature-detect}
\end{table}
In the \tool{} group, we integrated all $11$ features into the system. Participants reported that they noticed these features and were free to assign different priorities to each.
Thus, they take more time observing auxiliary information than the baseline group, which can be a reason for the significantly increased total labeling time.

According to the participants, the improved system provided features in advance and thus broadened their perspectives and encouraged further elicitation.
It also revealed subtle differences challenging or impossible to discern with the naked eye, aligned with \textit{C2 - Overlooking trajectory details}.
Though participants tended to set lower priority to some of the features, these features were crucial, especially in cases where their primary criteria exhibited limited variation and participants needed to turn to more features for guidance.
For example, one user stated that he would rely on lower-priority features like ``Power Usage'' when the primary criteria (\eg ``Collision'', ``Trajectory Smoothness'') cannot distinguish the trajectories.

\subsubsection{Other comments and suggestions}
\label{sec:results-comments-suggestions}
Participants provided various comments and suggestions regarding the system and the labeling process.

In terms of the labeling order and process, participants encountered certain pairs that were challenging to compare, possibly those where \textit{User Familiarity}, \textit{Pair disagreement} or \textit{Cluster Disagreement} dominated other metrics in \Cref{tab:ranking-metrics},
and some easier pairs, possibly those where \textit{Cluster Coverage} or \textit{Pair Similarity} dominated other metrics in \Cref{tab:ranking-metrics}.
Trajectory pairs with different difficulty levels are interspersed in the labeling process.
Therefore, the participants found the labeling process's workload more manageable and experienced increased engagement in the task.
Several participants mentioned that they would feel more rewarded if they could see the improvements in the robot's performance during the labeling process.

From the features and keyframes perspective,
some participants recommended that the system should always highlight unobservable features and provide more detailed explanations or definitions of the trajectory features,
particularly for those unfamiliar with the robotic task.
Moreover, participants wished to view the complete list of feature distributions and keyframes and select the ones to display and compare.
Some participants suggested that the system could provide more detailed written descriptions instead of letting the host explain the trajectory features orally.
One participant suggested that the system should provide a way to quantify the collision's severity since some collisions do not affect the safety of the object and the robot; in contrast, some collisions may cause severe damage.
A possible way is to use the collision force to quantify the severity of each collision.

From the workload and engagement perspective, although more engaged using \tool{}, many participants still feel overwhelmed by the number of trajectory pairs for the labeling workload and hope to have fewer pairs to label.
One insightful suggestion is leveraging gamification to make the labeling process more engaging by providing more variety in the task scenario presented.

%% file: sections/discussion.tex
\section{Discussion}
\label{sec:discussion}

Our user study shows that the participants in the \tool{} condition have significantly higher labeling consistency than those in the baseline condition.
Although the total labeling time of the \tool{} group is significantly longer than that of the baseline group,
the participants in the \tool{} group find it significantly easier to form comparison criteria and maintain engagement than those in the baseline group.
They also consider \tool{} helpful and easy to use.
According to participant ratings, our proposed design requirements of \tool{} help tackle the corresponding challenges identified in the formative study.
In the following subsections, we discuss the potential research implications derived from the participants' feedback, the room for improvement of each component, the generalizability of the proposed tool \tool{}, and the possible limitations of this work.

\subsection{Implications for Future Research}

\subsubsection{Support for sense-making and preference elicitation}
\label{sec:discussion-features-keyframes}

Our data-set \dataset{} augments conventional robot task data with descriptive statistics of a wide variety of trajectory features and keyframes of critical moments in task videos. We leverage such information to cluster robot trajectories, prompt labeling arrangement, and facilitate comparison on the \tool{} interface.
The results (\Cref{sec:results-c1,sec:results-c2,sec:results-design-requirements})
show that these features and keyframes can help users better understand complex trajectories, establish preference criteria, and make decisions.
Through the keyframes, users can visually identify the differences in a specific feature between the trajectories through a quick glance at the related video segment.
For example, users can compare the height at which the robot drop the object by the corresponding juxtaposed frames from the two videos.
With the plots of feature value distributions, users can determine the priority of the features.

Due to the space limitation, in our current design, \tool{} chooses to present the distribution of a feature on the labeling interface if its value in any of the given two trajectories is considered an outlier.
The assumption is that such features are distinctive and worthy of attention.
Another possible way to select features for display is to let users specify features of interest, as some participants mentioned in the interview.
However, users may not know what features matter to them, especially at the beginning, or their judgment is biased.
In other words, both the adaptive and the adaptable approaches have their own advantages and disadvantages.
Future work may further explore the usability and user-friendliness of a mixed-initiative approach \cite{buntSupportingInterfaceCustomization2007}, 
ensuring user agency while trying to mitigate individual biases. 
Overall, these features serve as an abstraction of the trajectories to boost comprehension. 
We only employ familiar basic charts and keyframes to illustrate the features in this work. Future research can explore alternative means to present such auxiliary information on the side or directly overlay on top of a task trajectory.
For example, a potential direction can be utilizing situated visualization in augmented reality (AR) \cite{chandanLearningVisualizationPolicies2023a} for preference elicitation of trajectories.

\subsubsection{Overconfidence in self-assessments}
\label{sec:discussion-overconfidence}

Our study finds that the metrics related to participants' self-assessments ({C3-1} [easy], {Confidence}, {C1-2} [feature coverage], {C1-3} [feature distribution], and {C2-1} [robotic knowledge]) did not show significant differences between the \tool{} and the baseline groups.
However, we observe a significant improvement in the {Consistency} metric of the \tool{} group.
Moreover, during the interviews, many baseline group participants admitted that they failed to notice some important features (\Cref{sec:results-interview-features-comparison-criteria}).
These results suggest that the participants may be overconfident in their preference labels, known as \textit{Dunning-Kruger effect} \cite{kruger1999unskilled}, even though most claimed to be unfamiliar with robotics.
This finding aligns with previous studies \cite{gadirajuUsingWorkerSelfAssessments2017, drawsChecklistCombatCognitive2021} that have identified the overconfidence bias in crowdsourcing tasks.
In such tasks, labelers with low ability tend to overestimate their performance, leading to {insufficient observations and inconsistent labels.}
This insight highlights the importance of assisting labelers in calibrating their confidence.
Possible methods include but are not limited to 
training the non-expert labelers \cite{vanderstappenGuidelinesDesigningHumanintheLoop2021} to improve their expertise,
increase labelers' awareness of cognitive biases by providing trial-by-trial feedback \cite{haddaraImpactFeedbackPerceptual2022}, 
and introducing labeler collaboration in the preference elicitation process \cite{warmathImpactSharedFinancial2019}
Future research can investigate the efficacy of various confidence calibration mechanisms and the consequent effect on label quality. 

\subsubsection{Design of prompting mechanisms}
\label{sec:discussion-prompting-order}

\tool{} tries to balance the model's information collection and the user's experience by considering the
similarities between trajectories,
disagreements among labelers, and individual familiarity with the trajectory pair when dynamically adjusting the comparison order.
The results from the \tool{} group suggest that presenting trajectory pairs that cover more distinct cases at the initial stage can improve their ability to observe a wide variety of details and establish relatively reliable criteria
(\Cref{sec:results-design-requirements}).
Additionally, according to the interview, 
the \tool's adaptive prompting order helps users stay engaged in the task and promotes a positive user experience.
Nevertheless, how each component (\eg pair similarity, user familiarity, and disagreements) impacts the performance and experience is yet to be assessed. Future work can conduct ablation studies to determine the effect of individual components, comparing the effectiveness of the robot reward models trained on the labeling results.
Furthermore, future studies can experiment with other factors and mechanisms to achieve an optimal balance between the model's uncertainty, users' experience, and labeling quality.
For example, inspired by active learning \cite{akrourAPRILActivePreference2012a, holladayActiveComparisonBased2016a, shinBenchmarksAlgorithmsOffline2022}, researchers could add a model informative dimension to determine the priority of comparative pairs.
They may also incorporate annotation curricula \cite{leeAnnotationCurriculaImplicitly2022} or gamification \cite{sevastjanovaritaQuestionCombGamificationApproach2021} in the labeling process to reduce labelers' learning curve and improve their engagement.

\subsection{Generalizability of \tool{}}
\label{sec:discussion-generalize}

Our study demonstrates the effectiveness of our pairwise preference labeling system, \tool, in improving the consistency of human feedback in the pick-and-place task. 
This subsection discusses the generalizability of \tool{}. 

\subsubsection{Generalizing to other tasks}
\label{sec:discussion-generalize-tasks}

\tool{} can also be generalized to other robot arm manipulation tasks besides the pick-and-place task.
In the formative study, we interviewed the participants about the issues they may encounter in labeling other robot tasks. 
They reflected that the three challenges identified in \Cref{sec:formative-study-challenges} exist regardless of the tasks.
They further commented that the criteria and trajectory features of interests are likely to be similar across different robot arm manipulation tasks, but the priority of individual criteria may vary in different scenarios.
That is to say, the criteria, features, challenges, and design requirements we summarized from the formative study are not limited to the pick-and-place task.
To tailor our design to other tasks, it is only necessary to redefine the list and the formulas of trajectory features and extract the corresponding keyframes according to the specific requirements of the new task.
For example, the \textit{Distance} feature can be defined as the distance of the pouring point to the cup in the pouring task.
Future work can further investigate the effectiveness of \tool{} adapting to other robot arm manipulation tasks. For example, how well \tool{} could be adapted to more complex tasks involving multiple objects, multiple parties, or multiple steps, such as the kitchen tasks in \cite{guptaRelayPolicyLearning2019}.

\subsubsection{Generalizing to other feedback types}
\label{sec:discussion-generalize-feedback}

Since the highlights of our design for \tool{} are the augmentation of the features and prompting strategies, it is flexible enough to accommodate various types of human feedback, including but not limited to
pairwise comparison \cite{kuhlmanEvaluatingPreferenceCollection2019a, qianLearningUserPreferences2015, glickmanAdaptivePairedComparison2005}, 
multiple ranking \cite{brownExtrapolatingSuboptimalDemonstrations2019, brownSafeImitationLearning2020, zhuPrincipledReinforcementLearning2023a, myersLearningMultimodalRewards2022a}, 
rating \cite{carteretteHereThere2008}, 
\etc, as long as the input data are consistent. 
The dataset \dataset{} and the ranking metrics in \Cref{tab:ranking-metrics} are prepared for individual trajectories on the server side, regardless of how the trajectories are presented and annotated on the client side. 
To generalize \tool{} to other feedback types, we only need to adjust the user interface and the prompting strategies accordingly.
For example, we can modify the user interface for rating feedback to display one trajectory video at a time with a rating scale. 
The prompting strategies can be revised to ask users to rate the trajectories with the most distinct features at the initial stage and then proceed to the trajectories with less familiarity and more disagreements.
Future work can further investigate the effectiveness of \tool{} adapting to other feedback types for learning approaches requiring different kinds of human input such as \cite{pmlr-v164-wilde22a}.

\subsection{Limitations and Future Work}
\label{sec:discussion-limitations-future-work}

Our study has several limitations. 
First, similar to most existing research on robot-human alignment \cite{ouyangTrainingLanguageModels2022,casperOpenProblemsFundamental2023}, our study is limited by the representativeness of the participants and data.
Although we recruited participants from different backgrounds, including students, researchers, and engineers, we did not have large samples of each group of stakeholders.
Also, we did not cover all age groups
and all domain expertise groups in robotics or data labeling systems.
Future works can compare the performance and experiences of different user groups to gain further insights.
Second, because it can be hard to externalize one's internal thinking and reasoning, the criteria and features we summarized from the formative study may not be a complete and fully accurate representation of human values.
We may introduce additional sensors \cite{zhangSelfAnnotationMethodsAligning2023} or design algorithms \cite{bobuFeatureExpansiveReward2021} to detect implicit preference in human feedback in future work to solve this limitation.
Third, due to the limited number of participants and their labeled data, we cannot evaluate the data quality by training and evaluating the reward model.
We plan to conduct large-scale crowd-sourcing studies using \tool{} to 
collect more preference data, evaluate the reward model by training and testing the model on the collected data, and adopt the latest robot task learning algorithms \cite{shinBenchmarksAlgorithmsOffline2022} to see the improvement of robot manipulation.
Fourth, to ensure the fairness of the between-subjects study, we kept the pool of trajectories for labeling the same for the two conditions and had both groups annotate the entire pool.
Hence, we only included $30$ trajectories in the pool -- a small number compared to many existing robot trajectory datasets -- so that the length of the study sessions was reasonable.
Also, we did not use any selection strategies to reduce the number of trajectory pairs to be labeled.
Future work can incorporate active learning \cite{akrourAPRILActivePreference2012a}, or semi-automatic labeling \cite{degregorioSemiautomaticLabelingDeep2020} approaches to reduce the users' workload when more preference data need to be collected over a bigger trajectory pool to train the reward model and evaluate the improvement of the robot task learning.

%% file: sections/conclusion.tex
\section{Conclusion}
\label{conclusion}

This paper presents \tool{}, a feature-augmented system for robot trajectory preference labeling.
\tool{} is designed to help users establish their criteria and compare the trajectories.
Through a formative study, we identified the criteria and features that users care about when labeling the trajectories and the challenges they face when labeling the trajectories.
We then derive the design requirements for \tool{}, generate a robot arm dataset, and build a web application with dynamic prompting, adaptive display of features and keyframes, and staged feedback.
We conducted a comprehensive user study to evaluate the effectiveness of \tool{} in improving the consistency of human labelers.
The results show that \tool{} can significantly improve human labelers' consistency and help labelers in more engaged preference elicitation.
The results also show that \tool{} can help users establish the criteria and compare the trajectories with improved user engagement.
We also discuss the generalizability and limitation of \tool{} and provide design considerations for future work in improving the user experience, incorporating algorithmic assistance, and improving the preference dataset for robot learning.

%% file: sections/acknowledgement.tex
\begin{acks}
    Many thanks to the anonymous reviewers for their insightful suggestions. 
    We thank all the participants for taking part in the studies and for sharing their valuable feedback.
    This research was supported in part by the InnoHK funding launched by the Innovation and Technology Commission, Hong Kong SAR and the HKUST \& HKPC Joint Laboratory grant HKPC22EG01-A.
\end{acks}

%% file: sections/appendix/feature-definitions.tex
\section{\dataset{} Feature Definitions}
\label{sec:feature-formula-definitions}

\input{sections/floats/tab-feature-keyframe-formula.tex}

\Cref{tab:feature-keyframe-formula} shows the formula for each feature as a time series and as a scalar in a trajectory complementary to \Cref{tab:feature-keyframe-measure}.
We stack all feature time series for each criterion to form a criterion vector series:
\begin{align*}
    \mathbf{safety}_i(t)
    = [ & \quad num\_collision_i(t), dis\_to\_left_i(t),  &  &    \\
        & \quad dis\_to\_right_i(t), dis\_to\_front_i(t), &  &    \\
        & \quad dis\_to\_back_i(t),  dis\_to\_table_i(t), &  &    \\
        & \quad contact\_force_i(t)                       &  & ];
\end{align*}
\begin{align*}
    \mathbf{efficiency}_i(t)
    = [ & \quad speed_i(t), eef\_pos_i(t),             &  &    \\
        & \quad can\_pos_i(t), pseudo\_cost_i(t)       &  & ]; \\
    \mathbf{task\_quality}_i(t)
    = [ & \quad speed\_smoothness_i(t),                &  &    \\
        & \quad trajectory\_smoothness_i(t),           &  &    \\
        & \quad orientation_i(t), grasp\_position_i(t) &  & ];
\end{align*}
which represent three criteria, \textit{Safety}, \textit{Efficiency}, and \textit{Task Quality}, for trajectory $i$ at $t$-th step.

%% file: sections/floats/tab-feature-keyframe-formula.tex
\begin{table*}
    \centering
    \setlength{\colonewidth}{\dimexpr .13\textwidth - 2\tabcolsep}
    \setlength{\coltwowidth}{\dimexpr .13\textwidth - 2\tabcolsep}
    \setlength{\colthreewidth}{\dimexpr .73\textwidth - 2\tabcolsep}
    \caption{Formula definitions for each feature as a time series and as a scalar in a trajectory. In the formulas, $i$ represents the trajectory, $s_i$ denotes the total number of steps (\ie the number of states) in trajectory $i$ and $t \in \{0, 1, \ldots, s_i\}$ denotes the step-index in the trajectory series.}
    \begin{tabular}{m{\colonewidth}m{\coltwowidth}m{\colthreewidth}}
        \toprule
        \textbf{Feature}
         &
        \multicolumn{2}{c}{\textbf{Formula}}
        \\
        \midrule
        \multicolumn{3}{c}{Safety}
        \\
        \midrule
        \multirow{2}{*}{Collision}
         &
        \multirow{1}{*}{Time Series}
         &
        $num\_collisions_i(t)$
        \\
        \cmidrule{2-3}
         &
        \multirow{1}{*}{Scalar}
         &
        $max\_num\_collisions_i = \max\{collision_i(t)\}_{t=0}^{s_i}$
        \\
        \cmidrule{1-3}
        \multirow{2}{*}{\parbox{\colonewidth}{Distance (to \\table edges)}}
         &
        \multirow{1}{*}{Time Series}
         &
        $dis\_to\_left_i(t)$, $dis\_to\_right_i(t)$, $dis\_to\_front_i(t)$, and $dis\_to\_back_i(t)$.
        \\
        \cmidrule{2-3}
         &
        \multirow{1}{*}{Scalar}
         &
        $min\_dis\_to\_edge_i = \min\left\{dis\_to\_left_i(t), dis\_to\_right_i(t), dis\_to\_front_i(t),   dis\_to\_back_i(t)\right\}_{t=0}^{s_i}$
        \\
        \cmidrule{1-3}
        \multirow{2}{*}{\parbox{\colonewidth}{Distance (to \\table surface)}}
         &
        {Time Series}
         &
        $dis\_to\_table_i(t)$
        \\
        \cmidrule{2-3}
         &
        \multirow{1}{*}{Scalar}
         &
        $max\_height\_to\_table_i = \max\left\{dis\_to\_table_i(t)\right\}_{t=0}^{s_i}$
        \\
        \cmidrule{1-3}
        \multirow{2}{*}{Contact force}
         &
        {Time Series}
         &
        $eef\_force_i(t)$
        \\
        \cmidrule{2-3}
         &
        \multirow{1}{*}{Scalar}
         &
        $max\_eef\_force_i = \max\{eef\_force_i(t)\}_{t=0}^{s_i}$
        \\
        \midrule
        \multicolumn{3}{c}{Efficiency}
        \\
        \midrule
        \multirow{2}{*}{Speed}
         &
        {Time Series}
         &
        $speed_i(t)$
        \\
        \cmidrule{2-3}
         &
        {Scalar}
         &
        $avg\_speed_i = \frac{1}{s_i}\sum_{t=0}^{s_i} speed_i(t)$
        \\
        \cmidrule{1-3}
        \multirow{7}{*}{Path Length}
         &
        {Time Series (1)}
         &
        $eef\_pos_i(t)$
        \\
        \cmidrule{2-3}
         &
        {Time Series (2)}
         &
        $can\_pos_i(t)$
        \\
        \cmidrule{2-3}
         &
        {Scalar (1)}
         &
        $reach\_length_i   = \sum_{t=1}^{s1_i} \lvert{eef\_pos_i(t) - eef\_pos_i(t-1)}\rvert$
        \\
        \cmidrule{2-3}
         &
        \multirow{1}{*}{Scalar (2)}
         &
        $grasp\_length_i   = \sum_{t=s1_i + 1}^{s2_i} \lvert{eef\_pos_i(t) - eef\_pos_i(t-1)}\rvert$
        \\
        \cmidrule{2-3}
         &
        {Scalar (3)}
         &
        $grasp\_length_i = \sum_{t=s1_i + 1}^{s3_i} \lvert{eef\_pos_i(t) - eef\_pos_i(t-1)}\rvert$
        \\
        \cmidrule{1-3}
        \multirow{4}{*}{Time}
         &
        {Scalar}
         &
        $total\_time_i(t) = \frac{s_i}{fps}$
        \\
        \cmidrule{2-3}
         &
        {Keyframe (1)}
         &
        The step $s1_i$ that the end-effector gripped the can when picking up. (\eg \Cref{fig:pick_up_point})
        \\
        \cmidrule{2-3}
         &
        {Keyframe (2)}
         &
        The step $s2_i$ that the end-effector releases the can when placing. (\eg \Cref{fig:release_point})
        \\
        \cmidrule{1-3}
        \multirow{3}{*}{Power Usage}
         &
        {Time Series}
         &
        $pseudo\_cost_i(t) = \sum_{\tau = 0}^{t} \sum_{j=1}^{dof} |q_j(\tau)|$
        \\
        \cmidrule{2-3}
         &
        {Scalar}
         &
        $pseudo\_cost_i = \sum_{\tau = 0}^{s_i} \sum_{j=1}^{dof} |q_j(\tau)|$
        \\
        \midrule
        \multicolumn{3}{c}{Task Quality}
        \\
        \midrule
        \multirow{3}{*}{\parbox{\colonewidth}{Speed        \\Smoothness}}
         &
        \multirow{1}{*}{Time Series}
         &
        $speed\_smoothness_i(t) = \frac{1}{t}\sum_{\tau = 0}^{t} \sqrt{\mathbf{a}(\tau)^2}$
        \\
        \cmidrule{2-3}
         &
        \multirow{1}{*}{Scalar}
         &
        $speed\_smoothness_i   = \frac{1}{s_i}\sum_{\tau = 0}^{s_i} \sqrt{\mathbf{a}(\tau)^2}$
        \\
        \cmidrule{1-3}
        \multirow{3}{*}{\parbox{\colonewidth}{Trajectory   \\Smoothness}}
         &
        \multirow{1}{*}{Time Series}
         &
        $trajectory\_smoothness_i(t) = \frac{1}{t}\sum_{\tau = 0}^{t}   \arccos{\frac{\mathbf{x}(\tau) \cdot \mathbf{x}(\tau+1)}{\lvert \mathbf{x}(\tau)\rvert \lvert\mathbf{x}(\tau + 1)\rvert}}$
        \\
        \cmidrule{2-3}
         &
        \multirow{1}{*}{Scalar}
         &
        $trajectory\_smoothness_i(t) = \frac{1}{s_i}\sum_{\tau = 0}^{s_i}\arccos{\frac{\mathbf{x}(\tau) \cdot \mathbf{x}(\tau+1)}{\lvert   \mathbf{x}(\tau)\rvert \lvert\mathbf{x}(\tau + 1)\rvert}}$
        \\
        \cmidrule{1-3}
        \multirow{1}{*}{Orientation}
         &
        \multirow{1}{*}{Time Series}
         &
        $orientation_i(t) = eef\_ori\_mat_i(t)^{-1}*can\_ori\_mat_i(t)$
        \\
        \cmidrule{1-3}
        {Grasp Position}
         &
        {Time Series}
         &
        $grasp\_position_i(t)   = can\_pos_i(t) - eef\_pos_i(t)$
        \\
        \bottomrule
    \end{tabular}
    \label{tab:feature-keyframe-formula}
\end{table*}